\documentclass[amsmath,floatfix,nofootinbib,notitlepage,prd,superscriptaddress,twocolumn]{revtex4-2}

\usepackage[T1]{fontenc}
\usepackage[usenames,dvipsnames]{xcolor}
\usepackage{aas_macros,graphicx,hyperref,orcidlink}
\usepackage{amsfonts,amssymb}
\usepackage{float}
\usepackage{multirow}
\usepackage{mathrsfs}
\usepackage[normalem]{ulem}
\allowdisplaybreaks
\numberwithin{equation}{section}

\makeatletter
\renewcommand*{\@textcolor}[3]{%
  \protect\leavevmode
  \begingroup
    \color#1{#2}#3%
  \endgroup
}
\makeatother

\newcommand{\br}[1]{\left[#1\right]}

\newcommand{\pa}[1]{\left(#1\right)}

\begin{document}

\title{Light Propagation Prescriptions for Black Hole Movies}

\author{Daniel Rojas-Paternina\,\orcidlink{0009-0004-4913-8710}}
\affiliation{Department of Physics, Universidad Nacional de Colombia, Bogot\'a, Colombia}
\email{drojasp@unal.edu.co}

\author{Alejandro C\'ardenas-Avenda\~no\,\orcidlink{0000-0001-9528-1826}}
\affiliation{Department of Physics, Wake Forest University, Winston-Salem, North Carolina 27109, USA}
\email{cardenas@wfu.edu}

\begin{abstract}
The spatiotemporal content of a black-hole movie is set jointly by source variability and by the distribution of light-travel times across the image. In the slow-light prescription, an image evaluated at fixed observer time contains photons emitted at different source times, whereas in fast light all rays sample a single source emission time. In this work we compare these light-propagation prescriptions through the lensing-band structure of Kerr geodesic delays in a controlled semi-analytic setting. For a given emitting geometry, black-hole spin, and observer inclination, we show how the coordinate-time delay distributions of Kerr null geodesics, decomposed by image order across lensing bands, can be compared with the source correlation time to quantify differences between light-propagation prescriptions. We find that when the intrinsic variability timescale is comparable to, or shorter than, the relevant delay spread, the high-inclination mismatch between fast- and slow-light curves can reach several tens of percent. Motivated by this geometric structure, we introduce \emph{brisk light}, an intermediate prescription that compresses each lensing-band delay map to its dominant temporal interval rather than collapsing the full image to a single source time. The proposed methodology provides both a practical criterion for when slow light matters and an efficient route to black-hole movies that retain the leading temporal imprint of strong lensing, a regime of direct relevance for future space-based VLBI targeting photon-ring observables.
\end{abstract}

\maketitle

\vspace{-2em}

\section{Introduction}

The time dependence of a black-hole movie is not set by the source alone. It also depends on the distribution of null-geodesic travel times across the observer screen: even a single frame at fixed observer time is assembled from photons emitted at different source times. The observed movie therefore reflects a competition between intrinsic source variability and the propagation-induced delay structure of the image. Different assumptions about light propagation then lead to different prescriptions for computing black-hole movies. 

In the slow-light prescription, a single image frame at observer time $t_{\rm o}$ is assembled from rays with screen-dependent emission times $t_{\rm s}(\alpha,\beta)$. The resulting image is not a snapshot of the plasma at one instant, but a mixture of different source times whose support is set by the Kerr geodesic structure. This is the same strong-lensing structure that gives rise to the nested sequence of direct and indirect images associated with the photon ring, together with the characteristic time delays between them~\cite{Gralla:2019LensingKerr}. In contrast, the fast-light prescription ties all rays in a frame to the same source time. Fast light does not remove highly lensed trajectories when those are explicitly traced; rather, it erases their relative timing by collapsing the full delay field to a single emission snapshot.

The fast-light approximation is often not only the practical choice but, in some cases, the only computationally feasible approach. Slow light requires access to the source over the full temporal support sampled by a frame, and in general relativistic magnetohydrodynamics (GRMHD) post-processing this means closely spaced fluid snapshots, interpolation across multiple dumps, and substantially greater storage and throughput demands~\cite{Wong:2022PATOKA}. The Event Horizon Telescope (EHT) model libraries have therefore relied heavily on fast-light radiative transfer in model comparisons for both M87$^*$ and Sgr A$^*$~\cite{EventHorizonTelescope:2019M87V,EventHorizonTelescope:2022SgrAV,Wong:2022PATOKA}. This use is well motivated, but it also turns the central question into one of accuracy: when does a single source time represent the relevant delay support of the image?

Early GRMHD-based radiative-transfer studies already suggested that the fast-light approximation can be adequate for some near-horizon, subrelativistic accretion-flow images~\cite{Dexter:2010SgrASubmm}. A useful benchmark was later provided in Ref.~\cite{Bronzwaer:2018RAPTOR}: for a particular two-dimensional GRMHD setup, fast-light and slow-light integrated light curves differed by less than about five percent. Similar slow--fast comparisons have also been carried out in polarized GRRT calculations of magnetically arrested accretion flows~\cite{Moscibrodzka:2021CircularMAD}. 

Other time-domain analyses point to a more nuanced conclusion. Pattern-speed measurements in EHT-like movies explicitly identify the fast-light approximation as a possible source of uncertainty, and a high-spin, moderately inclined test model showed a non-negligible difference between fast and slow light~\cite{Conroy:2023RotationEHTMovies}. More recently, slow-light snapshots of the M87 jet-launching region found that fast and slow light can produce qualitatively different image morphologies when relativistic jet plasma is viewed close to the line of sight~\cite{Tsunetoe:2026SlowLightM87}. Photon-ring autocorrelations, two-point image-correlation functions, and related echo observables are more directly sensitive to the time ordering imposed by lensing~\cite{Hadar:2020PhotonAutocorr,Hadar:2023SpectroTemporal,Cardenas-Avendano:2024AbsenceSecondaryPeaks,Wong:2024LightEchoesVLBI,Bezdekova:2025CorrelationsMovies}. Thus, the fast-light approximation is not a yes-or-no prescription; it is controlled by the ratio of the source evolution time to the relevant propagation delay spread.

This distinction between light-propagation prescriptions is becoming observationally important. Space-based extensions of millimeter very-long-baseline interferometry (VLBI), including the Black Hole Explorer (BHEX), are designed to access the first photon ring and to exploit its strong-field geometry~\cite{Johnson:2024BHEXVision,Lupsasca:2024BHEXPhotonRing}. Analytical and semi-analytic accretion models also show that indirect images can carry measurable information even when they contribute only modestly to the total flux~\cite{Vincent:2022ThickDisks,Gralla:2020ObservableShape}. In such regimes, the relative delays between lensing orders are not a numerical detail; they are part of the observable.

In this work we study light-propagation prescriptions for black-hole movies in a controlled setting. We combine source movies generated by the time-dependent equatorial model \texttt{inoisy}~\cite{Lee:2021inoisy}, which represents the disk brightness as an inhomogeneous, anisotropic, time-dependent Gaussian random field with prescribed covariance, with ray tracing performed by \texttt{AART}~\cite{Cardenas-Avendano:2022AART}. To our knowledge, the \texttt{AART}+\texttt{inoisy} combination is currently the only publicly documented semi-analytic framework that provides controlled black-hole movies with tunable variability. The advantage of this setup is not that it uniquely separates propagation from emission---that separation is also standard in GRMHD post-processing---but that it provides a clean laboratory in which the source correlation time, the propagation prescription, and the lensing-band delay structure can be varied and interpreted separately.

Throughout this paper we follow the convention used in \texttt{AART}: lensing bands are annular regions on the observer screen associated with the equatorial-plane crossing order of Kerr geodesics, with $n=0$ denoting the direct image and $n\geq1$ denoting the indirect images~\cite{Cardenas-Avendano:2022AART,LensingBandsGeometric}. We refer to these higher-order indirect subimages as photon-ring images; in this terminology, the photon ring is the sequence of increasingly narrow lensing-order subimages that accumulates on the critical curve.

We show that, at high inclination, the fast-light and slow-light light curves can differ by substantially more than the few-percent level (e.g., with relative $L^1$ and $L^2$ differences of order $30$--$45\%$ at an observer inclination of $\theta_{\rm o}=60^\circ$). This discrepancy is explained by the band-resolved distribution of emission times and motivates an intermediate prescription that we call \emph{brisk light}. Brisk light keeps the dominant delay interval of each lensing band and clips only the low-probability tails of the delay distribution. It is therefore computationally less expensive than slow light, reducing the effective temporal support required for source interpolation, while retaining the leading temporal separation between direct and indirect emission that fast light erases.

Operationally, brisk light requires only the ray-traced emission-time map for each lensing band. For every band, one constructs the distribution of source times over the screen, identifies the modal highest-density interval enclosing a chosen probability mass $p$, and replaces the slow-light emission time by a clipped time: pixels whose emission times lie inside the interval are left unchanged, while pixels in the tails are mapped to the nearest interval boundary. The source movie is then evaluated at the same ray-traced source positions and with the same transfer factors used in slow light. The limit $p=0$ gives one representative time per lensing band, while $p=1$ recovers the discrete slow-light calculation; intermediate values retain the dominant bandwise delay support at reduced temporal cost.

The rest of the paper is organized as follows. Section~\ref{sec:setup} describes the ray tracing, source model, and diagnostics. Section~\ref{sec:temporal} analyzes the temporal support of the lensing bands. Section~\ref{sec:fastslow} compares fast and slow light and identifies the geometric origin of their differences. Section~\ref{sec:brisk} defines brisk light and benchmarks it against slow light. Section~\ref{sec:discussion} summarizes the implications. Numerical convergence is presented in Appendix~\ref{app:convergence}. We work in geometrized units with $G=c=1$, so all lengths and times are quoted in units of the black-hole mass $M$.

\section{Setup and diagnostics}
\label{sec:setup}

\subsection{Image construction}

We compute the Kerr transfer map with \texttt{AART}, which analytically traces null geodesics from the observer screen to an equatorial emitting plane and decomposes the image into lensing bands~\cite{Cardenas-Avendano:2022AART}. We label these bands by $n=0,1,2,\ldots$, where $n=0$ denotes the direct image and $n\geq1$ denotes successively higher-order lensed images.

For each screen point $(\alpha,\beta)$ in band $n$, the ray tracer returns the source event, in Boyer--Lindquist coordinates, $(t_{\rm s}^{(n)},r_{\rm s}^{(n)},\phi_{\rm s}^{(n)})=(t_{\rm s}^{(n)}(\alpha,\beta), r_{\rm s}^{(n)}(\alpha,\beta), \phi_{\rm s}^{(n)}(\alpha,\beta))$ together with the redshift factor $g_n(\alpha,\beta)=\nu_{\rm o}/\nu_{\rm s}$, where $\nu_{\rm o}$ and $\nu_{\rm s}$ are the photon frequencies measured by the observer and by the emitter, respectively. The redshift is determined by the photon four-momentum $k_\mu^{(n)}$ along the ray and by the observer and emitter four-velocities. We use a Keplerian-like equatorial velocity prescription and refer the reader to Ref.~\cite{Cardenas-Avendano:2022AART} for further details of the geodesic map, lensing-band construction, and source kinematics. 

The source-frame intensity is supplied by \texttt{inoisy}~\cite{Lee:2021inoisy}, which provides stochastic, nonaxisymmetric source movies with prescribed spatial and temporal correlations. The \texttt{AART}+\texttt{inoisy} pipeline therefore provides a semi-analytic route for generating controlled time-dependent equatorial black-hole source movies without running a GRMHD simulation. We denote the source-frame intensity by $I_{\rm{s}}=I_{\rm{s}}(t,r,\phi)$ and its temporal correlation scale by $\lambda_0$. In the simulations shown below, we take $\lambda_0$ to be inversely proportional to the Keplerian frequency (i.e., proportional to $r_{\rm{s}}^{3/2}$) as done in Ref.~\cite{Cardenas-Avendano:2022AART}. At a given radius, this timescale can be compared with the band-resolved delay widths reported in Sec.~\ref{sec:temporal}.

Slow-light ($X=\mathrm{s}$) provides the physically faithful description of light propagation by sampling the source at the corresponding retarded time along each Kerr geodesic, whereas fast-light ($X=\mathrm{f}$) evaluates all rays at a single source time $t_{\rm o}$. To retain part of the temporal structure of the slow-light delay map while reducing computational cost, below we introduce brisk-light ($X=\mathrm{b}$) as an intermediate prescription. The contribution of the $n$th lensing band to the observed image at observer time $t_{\rm o}$ is~\cite{Cardenas-Avendano:2022AART}
\begin{equation}
    I_{{\rm o},X}^{(n)}(t_{\rm o})
    =
    g_n^3\,
    I_{\rm s}
    \!\left[
        t_X^{(n)}(t_{\rm o}),
        r_{\rm s}^{(n)},
        \phi_{\rm s}^{(n)}
    \right].
    \label{eq:band_image_g3}
\end{equation}
The factor $g_n^3$ transfers source-frame specific intensity to observed specific intensity along that ray, $I_{\rm o}=g_n^3 I_{\rm s}$, and follows from the invariance of $I_\nu/\nu^3$. The redshift $g_n$ includes both gravitational and Doppler frequency shifts for the adopted emitter motion. All propagation prescriptions considered in this paper use the same geodesic map, source positions, and redshift factors; they differ only in the source time $t_X^{(n)}$ at which the movie is sampled. After subtracting the common observer-radius contribution (defined in Eq.~\ref{eq:renormalized_time}), the slow-light source time used in the movie can be written as $t_{\rm slow}^{(n)}(t_{\rm o},\alpha,\beta)=t_{\rm o}+\tilde t_{\rm s}^{(n)}(\alpha,\beta)$, whereas fast light uses $t_{\rm fast}(t_{\rm o})=t_{\rm o}$ for all pixels and all lensing bands. Thus, in the renormalized convention used below, fast light corresponds to the relative reference isochrone $\tilde t_{\rm s}=0$.

The total observed image is obtained by summing over the retained lensing bands,
\begin{equation}
    I_{{\rm o},X}(t_{\rm o},\alpha,\beta)
    =
    \sum_{n=0}^{n_{\rm max}}
    I_{{\rm o},X}^{(n)}(t_{\rm o},\alpha,\beta).
    \label{eq:total_observed_image}
\end{equation}
In slow light, $t_X^{(n)}$ is the full screen-dependent emission-time map returned by the ray tracing. In fast light, all rays in a frame are evaluated at a single source time. The brisk-light time map, defined in Sec.~\ref{sec:brisk}, modifies only the temporal argument of Eq.~\eqref{eq:band_image_g3}.

Unless explicitly noted otherwise, we work with \texttt{inoisy} movies that cover a total duration of $5000M$ on a uniform temporal grid of $8192$ frames, implying an intrinsic source cadence of $\Delta T\simeq 0.61M$. The spatial resolution of the \texttt{inoisy} simulations is $\Delta X\simeq 0.1M$. The black-hole movies are generated with an observer-frame cadence of $\Delta t_{\rm o}\simeq 1.22M$ on a screen with pixel spacing $\Delta x\simeq 0.1M$. Throughout this work, we fix the black-hole spin to $a/M=0.94$, consider observer inclinations of $\theta_{\rm o}=17^\circ$ and $60^\circ$, and retain the first three lensing bands, $n=0,1,2$.

\subsection{Light-curve diagnostics}

For each prescription $X$, we define the screen-integrated light curve
\begin{equation}
    \mathscr{L}_{X}(t_{\rm o})
    =
    \int
    I_{{\rm o},X}(t_{\rm o},\alpha,\beta)\,
    d\alpha\,d\beta ,
    \label{eq:light_curve_definition}
\end{equation}
with the integral evaluated numerically on the image grid. The slow-light light curve is denoted by $\mathscr{L}_{\rm s}(t)$ and is used as the reference curve unless otherwise stated. The subscript ``s'' therefore has two distinct uses: $I_{\rm s}$ denotes the source-frame intensity, while $\mathscr{L}_{\rm s}$ denotes the slow-light light curve.

For the comparisons, we rescale each light curve to a common mean value,
\begin{equation}
    \bar{\mathscr{L}}_{X}(t_j)
    =
    \frac{\mathscr{L}_{\rm target}}
    {\left\langle \mathscr{L}_{X}\right\rangle}
    \mathscr{L}_{X}(t_j),
    \,\,
    \left\langle \mathscr{L}_{X}\right\rangle
    =
    \frac{1}{N}\sum_{j=1}^{N}\mathscr{L}_{X}(t_j),
    \label{eq:mean_normalization}
\end{equation}
where $\mathscr{L}_{\rm target}=0.6$ in our implementation. This choice only fixes the vertical scale of the plotted light curves and has no effect on the relative-error metrics. The specific value is chosen so that the light curves are visually comparable to the total compact flux of Sgr A* in Jy. The signed pointwise relative difference between a reference curve $A$ and a comparison curve $B$ is then
\begin{equation}
    \Delta_{A,B}(t_j)
    =
    100\,
    \frac{
    \bar{\mathscr{L}}_{A}(t_j)
    -
    \bar{\mathscr{L}}_{B}(t_j)}
    {\bar{\mathscr{L}}_{A}(t_j)} .
    \label{eq:relative_difference}
\end{equation}
Positive values therefore indicate that the comparison curve lies below the reference curve. For aggregate comparisons, we subtract the mean and divide by the standard deviation of each light curve,
\begin{equation}
    \widetilde{\mathscr{L}}_{X}(t_j)
    =
    \frac{
    \mathscr{L}_{X}(t_j)
    -
    \left\langle \mathscr{L}_{X}\right\rangle}
    {\sigma_X},    \label{eq:standardized_light_curve}
\end{equation}
where 
\begin{equation}
    \sigma_X^2
    =
    \frac{1}{N}
    \sum_{j=1}^{N}
    \left[
    \mathscr{L}_{X}(t_j)
    -
    \left\langle \mathscr{L}_{X}\right\rangle
    \right]^2.
\end{equation}
Standardization isolates temporal structure from absolute normalization. For two standardized curves sampled on a common observer-time grid, we use these two standard norms
\begin{align}
    L^1(A,B)
    &=
    \frac{
    \int
    \left|
    \widetilde{\mathscr{L}}_{A}(t)
    -
    \widetilde{\mathscr{L}}_{B}(t)
    \right|\,dt}
    {
    \int
    \left|
    \widetilde{\mathscr{L}}_{A}(t)
    \right|\,dt},
    \label{eq:l1_def}
    \\
    L^2(A,B)
    &=
    \left[
    \frac{
    \int
    \left|
    \widetilde{\mathscr{L}}_{A}(t)
    -
    \widetilde{\mathscr{L}}_{B}(t)
    \right|^2\,dt}
    {
    \int
    \left|
    \widetilde{\mathscr{L}}_{A}(t)
    \right|^2\,dt}
    \right]^{1/2}.
    \label{eq:l2_def}
\end{align}
The integrals are evaluated numerically, with $A$ taken to be the reference prescription, usually slow light. These $L^1$ and $L^2$ distances are the primary light-curve diagnostics used below.

The numerical stability of these comparisons is checked in Appendix~\ref{app:convergence}, where we vary the screen resolution, the source cadence, and the ray-tracing cadence. These tests show that the reported propagation differences are stable once the image grid is sufficiently resolved, and that coarsening the source cadence mainly suppresses short-timescale variability rather than introducing a new propagation effect.

\section{Temporal structure of the lensing bands}
\label{sec:temporal}

For each lensing band, we study the distribution of source times over the screen. We remove the large observer-radius, $r_{\rm o}$, contribution by defining the renormalized emission time~\cite{Cardenas-Avendano:2022AART}
\begin{equation}
    \tilde t_{\rm s}^{(n)}(\alpha,\beta)
    =t_{\rm s}^{(n)}(\alpha,\beta)-\br{r_{\rm o}+2M\log\pa{\frac{r_{\rm o}}{M}}}.
    \label{eq:renormalized_time}
\end{equation}
Since the subtraction is an additive constant common to all rays, the relative structure of the delay map is unchanged. For the examples shown in this work, we have set $r_{\rm o}=1\times10^4\,M$ without loss of generality.

\begin{figure}[]
    \centering
    \includegraphics[width=\linewidth]{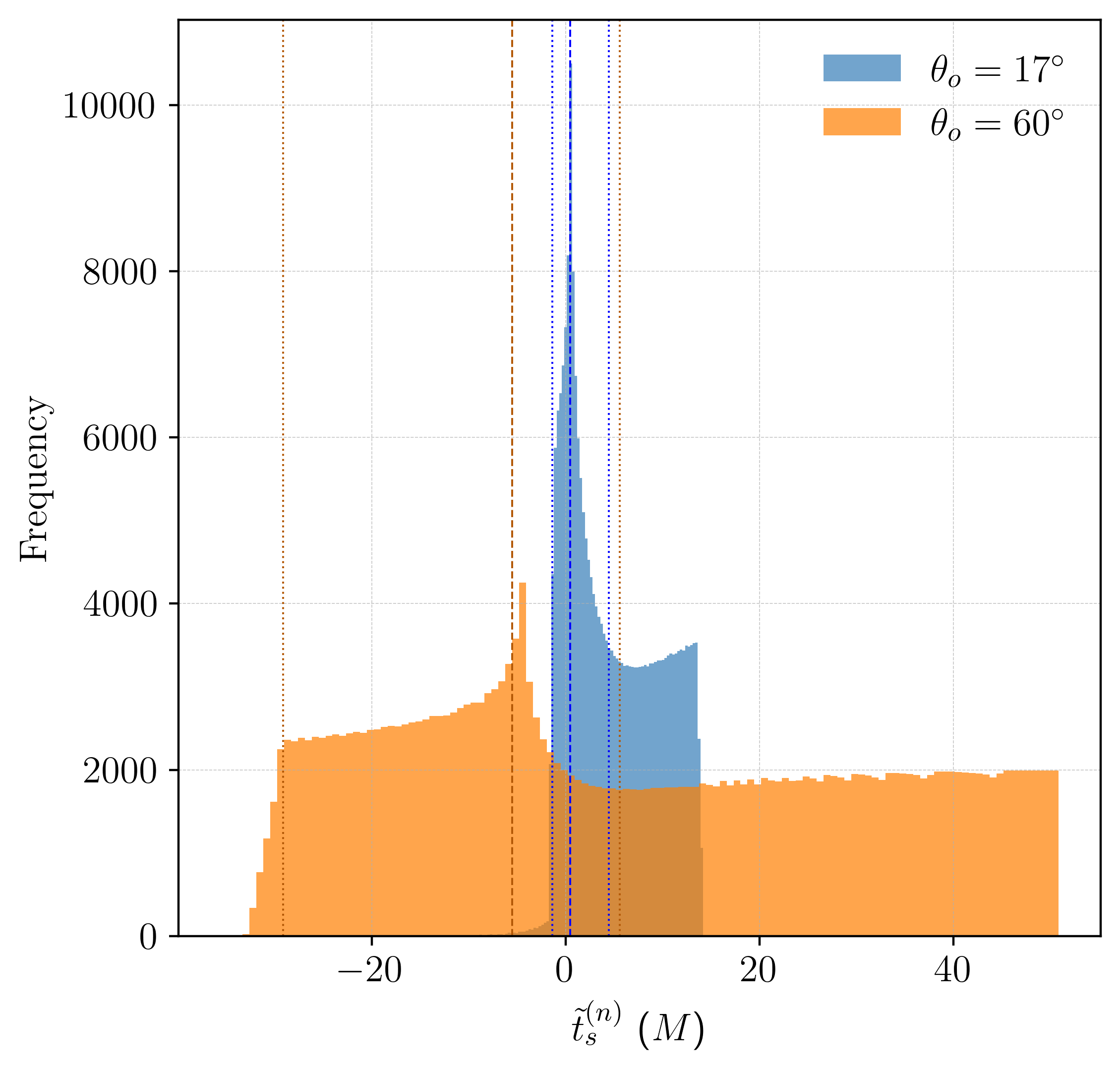}
    \caption{Histograms of the renormalized emission time $\tilde t_{\rm s}^{(0)}$ for direct-image pixels ($n=0$) at $\theta_{\rm o}=17^\circ$ and $60^\circ$. For each distribution, the vertical lines mark, from left to right, the lower endpoint, mode, and upper endpoint of the modal highest-density interval corresponding to $p=0.5$: $(-1.4,\,0.4,\,4.4)\,M$ for $\theta_{\rm o}=17^\circ$ and $(-29.2,\,-5.6,\,5.6)\,M$ for $\theta_{\rm o}=60^\circ$. The high-inclination case spans a wider range of source times, indicating that a single fast-light snapshot represents a smaller fraction of the available delay support.}
    \label{fig:Hist_Total}
\end{figure}

Figure~\ref{fig:Hist_Total} compares the $n=0$ distribution for two observer inclinations. The low-inclination distribution is narrow and sharply peaked, while the high-inclination distribution is broader and flatter, reflecting the larger range of source positions and photon paths visible to the observer. This difference already explains the qualitative behavior of fast light: near face-on, one representative source time is a good approximation to much of the direct image; at higher inclination, a single time discards more of the delay field.

\begin{figure}[]
    \centering
    \includegraphics[width=\linewidth]{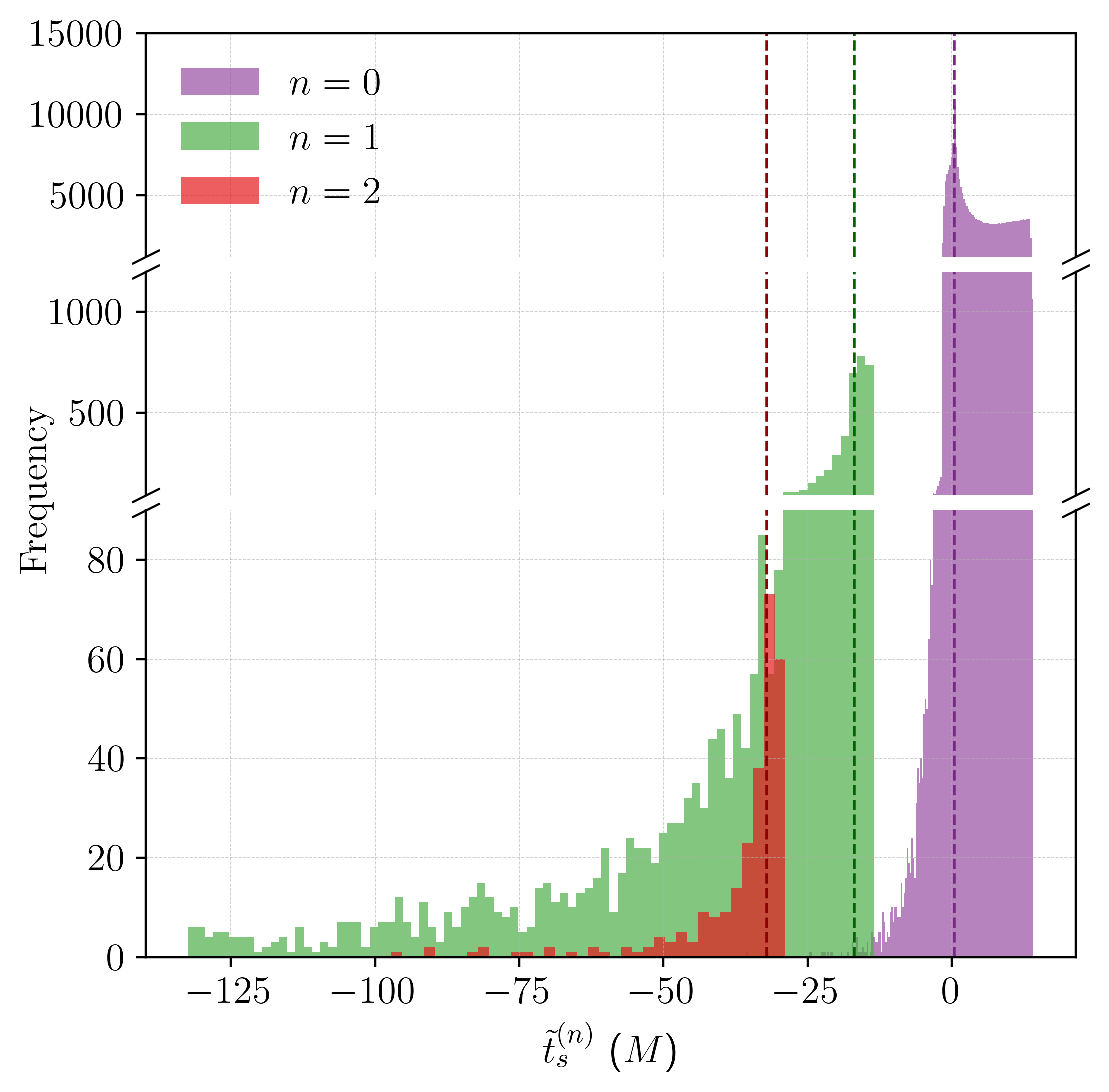}
    \caption{Band-resolved distributions of $\tilde t_{\rm s}^{(n)}$ at $\theta_{\rm o}=17^\circ$ for $n=0,1,2$. The direct image and the first two indirect images populate separated temporal windows, with modes at $\tilde t_{\rm s}^{(0)} = 0.4\,M$, $\tilde t_{\rm s}^{(1)} = -16.9\,M$, and $\tilde t_{\rm s}^{(2)} = -32.1\,M$ (vertical dashed lines). The broken vertical scale makes both the large direct-image population and the smaller higher-order populations visible.}
    \label{fig:Hist_17_log}
\end{figure}

The lensing-band decomposition makes the delay ordering explicit. Figure~\ref{fig:Hist_17_log} shows the first three lensing bands at $\theta_{\rm o}=17^\circ$. The direct image and the first two indirect images occupy distinct temporal windows because higher-order photons spend more time winding around the black hole before reaching the observer. These bands are therefore shifted to earlier source times, or equivalently to longer travel times. This band separation is the geometric origin of photon-ring echo delays~\cite{Gralla:2019LensingKerr,Hadar:2020PhotonAutocorr,Hadar:2023SpectroTemporal}.

Throughout the main analysis we fix the spin to $a/M=0.94$. Spin enters the problem through the Kerr transfer map: it changes the critical curve, the shapes and areas of the lensing bands, the redshift factors, and the coordinate-time delays accumulated by strongly bent rays. In a Keplerian-like source model it also changes the orbital frequencies assigned to material in the emitting plane. We choose a high-spin configuration as a strong-field stress test, where frame dragging and near-horizon delay structure are pronounced. Varying the spin would change the numerical delay widths and flux weights of the bands, but not the central methodological question addressed here, namely how a fixed delay map should be sampled for different light prescriptions. 

\section{Fast-light versus slow-light}
\label{sec:fastslow}

The temporal structure of the lensing bands turns the validity of fast light into a timescale comparison. If the source correlation time $\lambda_0$ is long compared with the relevant delay width $\sigma_n$, then the slow-light image samples nearly the same source phase across that band. If $\lambda_0\lesssim\sigma_n$, the image combines genuinely different source phases. 

The same conclusion follows from a local expansion. For one lensing band, let $\bar t_n$ be a representative source time, such as the mode or mean of the band delay distribution. At fixed ray-traced source position $(r_{\rm s}^{(n)},\phi_{\rm s}^{(n)})$, expanding the source intensity about $\bar t_n$ gives
\begin{equation}
    I_{\rm s}(t_{\rm s}^{(n)})
    =I_{\rm s}(\bar t_n)
    +\dot I_{\rm s}(\bar t_n)\bigl(t_{\rm s}^{(n)}-\bar t_n\bigr)+\cdots.
    \label{eq:expansion_about_representative_time}
\end{equation}
After summing over the screen, the correction to a single-time approximation is controlled by the moments of the delay distribution and by the time derivatives of the source. The fast-light approximation is therefore accurate when the dominant delay spread is small compared with the source variability time, and least reliable when the image samples a broad range of source times over which the source intensity changes appreciably.

\begin{figure}[]
    \centering
    \includegraphics[width=\linewidth]{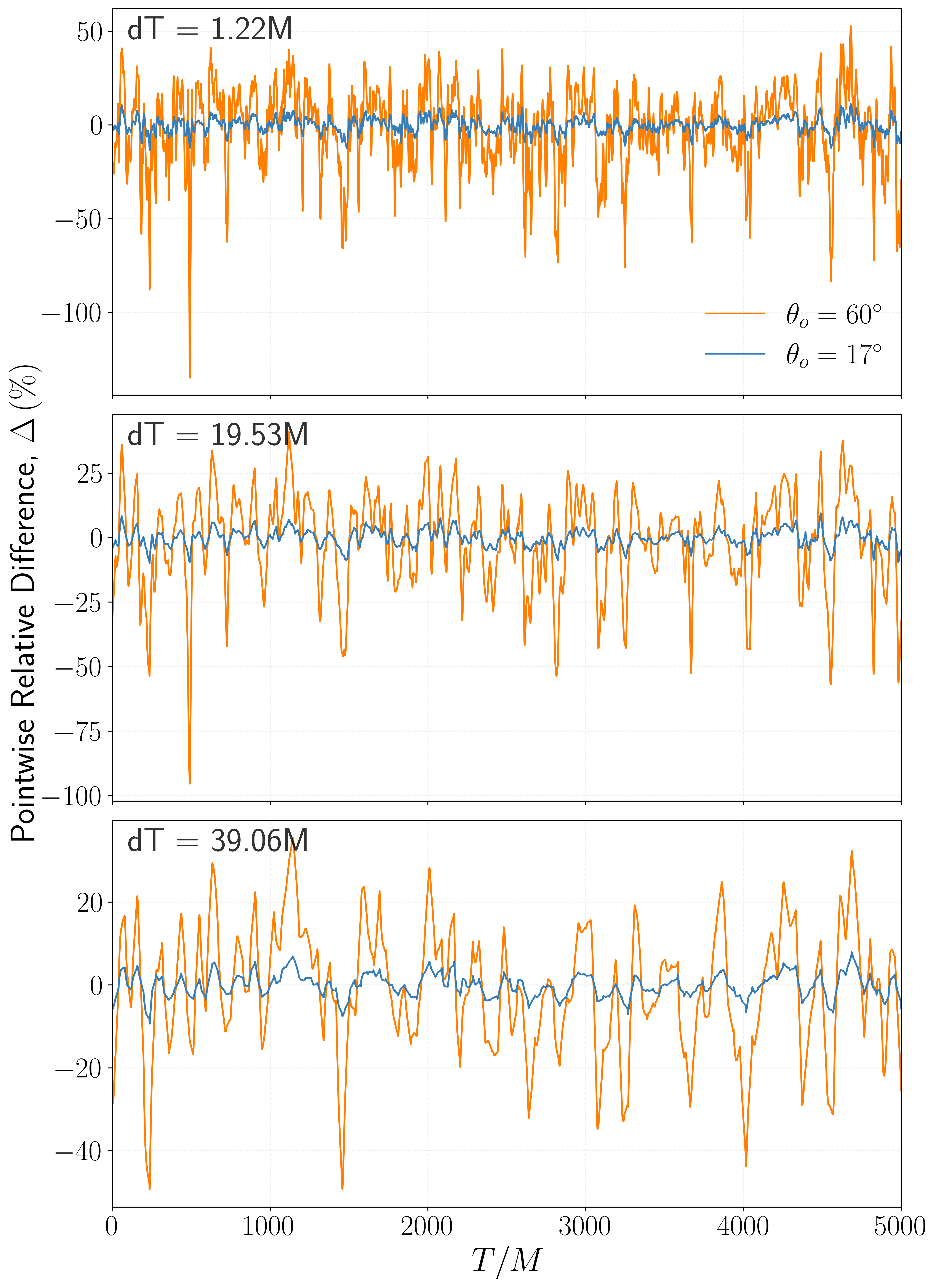}
    \caption{Pointwise relative difference $\Delta(t)$ between mean-normalized slow-light and fast-light light curves for several source cadences $\Delta T$. Each panel fixes the cadence and compares $\theta_{\rm o}=17^\circ$ (blue lines) with $60^\circ$ (orange lines). The high-inclination curve shows larger excursions, while coarsening the source movie suppresses the discrepancy by removing short-timescale variability.}
    \label{fig:ER_dt_60}
\end{figure}

\begin{figure}[]
    \centering
    \includegraphics[width=\linewidth]{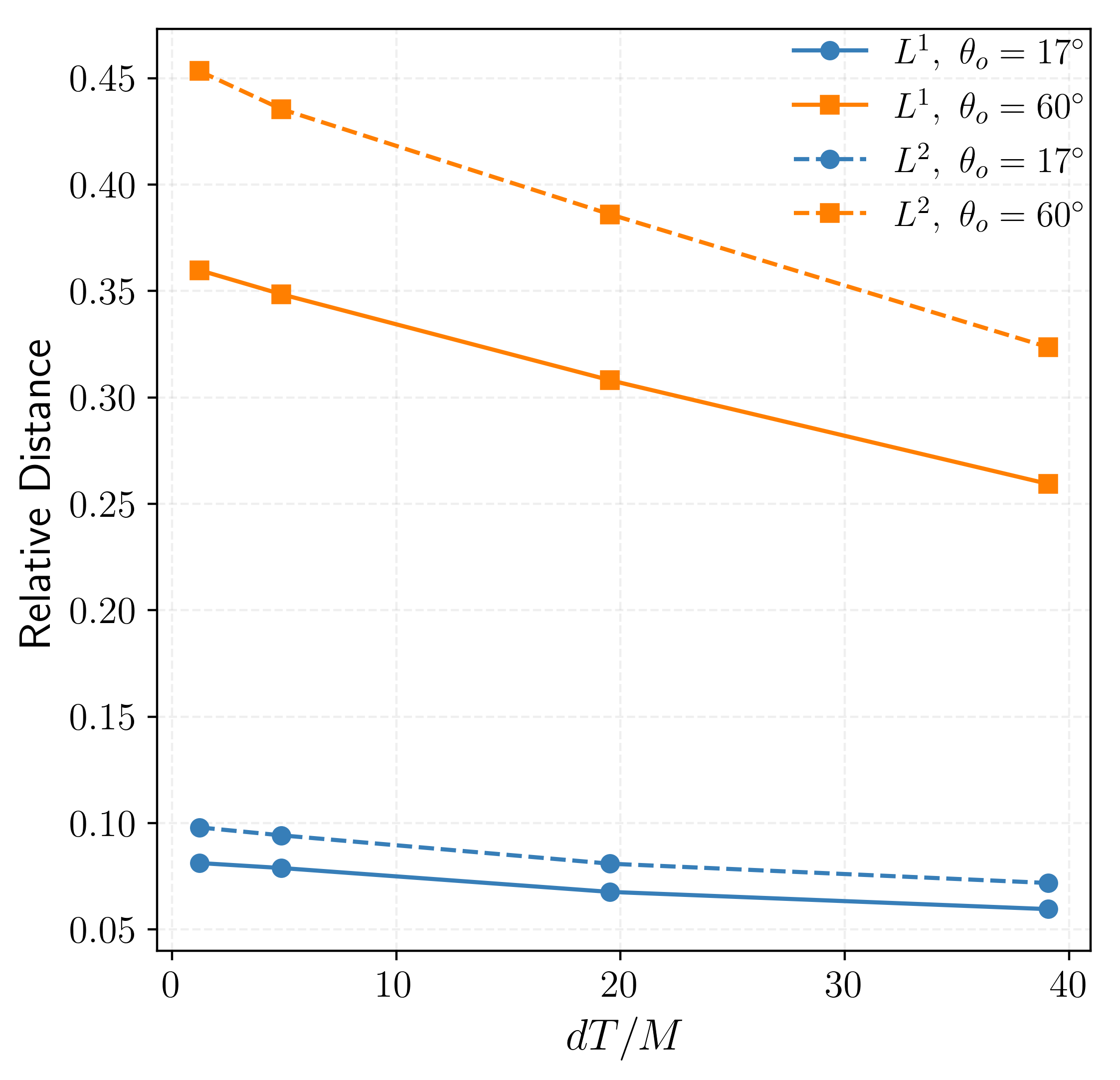}
    \caption{Relative $L^1$ and $L^2$ distances between standardized slow-light and fast-light light curves as functions of the source cadence $\Delta T$. The metrics confirm the trends visible in Fig.~\ref{fig:ER_dt_60}: fast light is much more accurate at low inclination (blue lines), and the difference between prescriptions decreases when the source variability is temporally under-resolved.}
    \label{fig:ErrL1L2_dT}
\end{figure}

Figures~\ref{fig:ER_dt_60} and~\ref{fig:ErrL1L2_dT} quantify this behavior. For the low-inclination case, $\theta_{\rm o}=17^\circ$ (blue lines), the relative difference remains small throughout the movie. For $\theta_{\rm o}=60^\circ$ (orange lines), the discrepancy is substantially larger, with signed pointwise deviations well above the few-percent level at the finest source cadence. For example, at $\Delta T=4.88\,M$, we find $L^1_{\rm slow,fast}=0.35$ and $L^2_{\rm slow,fast}=0.44$ at $\theta_{\rm o}=60^\circ$, compared with $L^1_{\rm slow,fast}=0.08$ and $L^2_{\rm slow,fast}=0.09$ at $\theta_{\rm o}=17^\circ$.

Increasing $\Delta T$ reduces the discrepancy because the high-frequency source variability is no longer resolved. In that limit, many different emission times correspond to nearly the same interpolated source state, so the distinction between fast and slow light is partially erased. For the cases shown in Fig.~\ref{fig:ErrL1L2_dT}, the relative $L^1$ and $L^2$ distances are larger at higher inclination, as expected from the broader delay spread.

Although the numerical values of the errors depend on the source model and on the details of the radiative-transfer implementation, the organizing criterion is geometric. Fast light works when the delay support that contributes appreciably to the image is narrow relative to the source correlation time. It significantly deteriorates when a single source snapshot cannot represent the range of emission times selected by the Kerr transfer map. 

This interpretation is consistent with Ref.~\cite{Bronzwaer:2018RAPTOR}, which found that fast-light and slow-light integrated light curves can differ by only a few percent for a particular two-dimensional GRMHD configuration viewed at low inclination. It is also consistent with the contrasting behavior found in recent slow-light general relativistic radiative transfer calculations of the M87 jet-launching region, where the emitting plasma becomes relativistic and is viewed nearly along the line of sight; in that regime, a single fast-light snapshot cannot represent the evolving material sampled along the light paths, and slow light produces qualitatively different jet morphologies~\cite{Tsunetoe:2026SlowLightM87}. It is also consistent with recent GRMHD correlation measurements showing that the slow-light two-point image correlation contains a lensing lobe near the expected indirect-image delay, while the corresponding fast-light lobe is shifted toward zero lag~\cite{Bezdekova:2025CorrelationsMovies}. Since these effects are tied to the relative timing of direct and indirect emission, we next isolate how the higher-order images enter the fast--slow comparison.

\subsection{The contribution of the higher-order images}

\begin{figure}[]
    \centering
    \includegraphics[width=\linewidth]{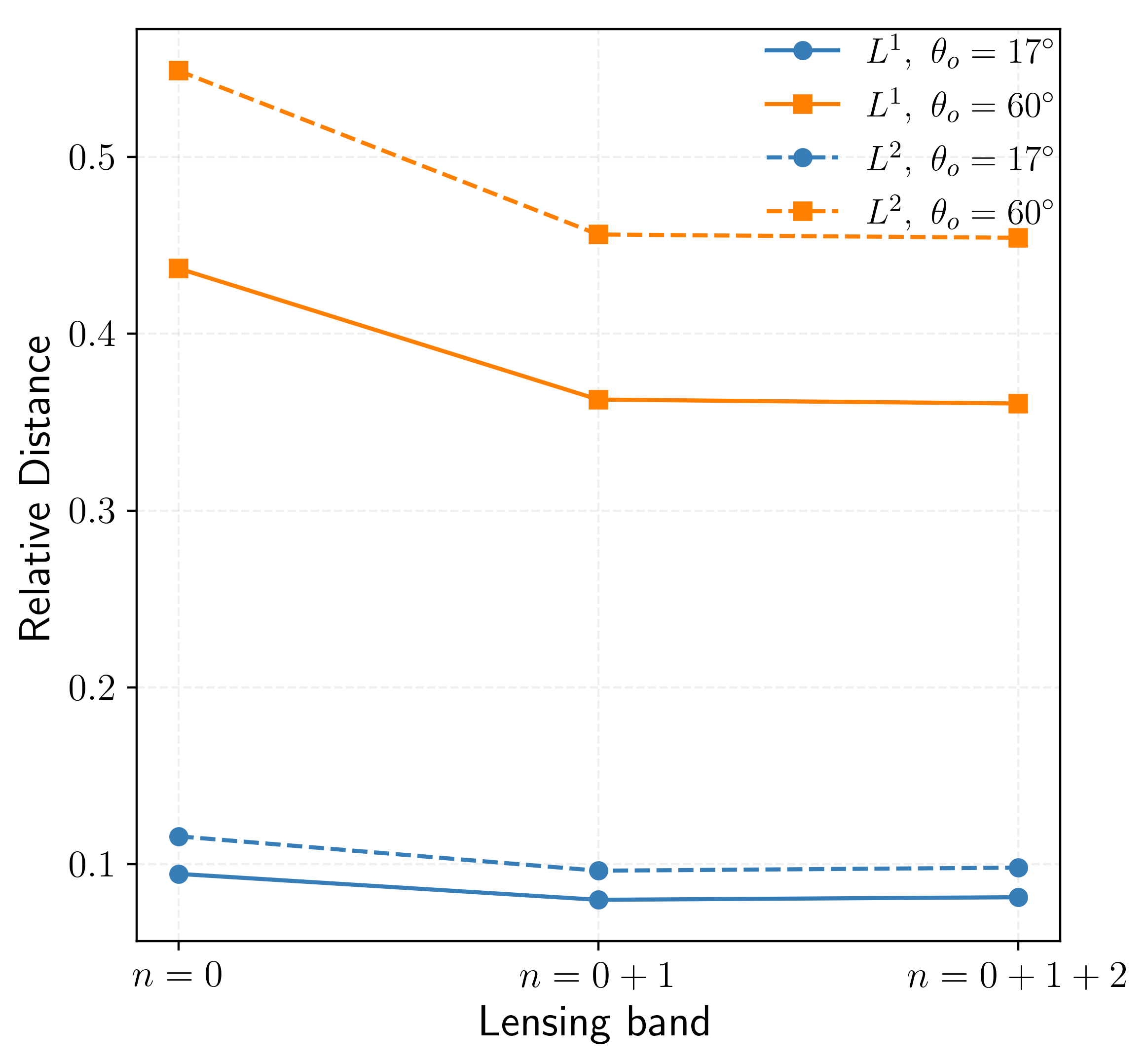}
    \caption{Relative $L^1$ and $L^2$ distances between standardized slow-light and fast-light light curves after truncating the image at different lensing orders. The fast--slow mismatch in this source model is not dominated by the highest retained order; it is already present in the direct image and changes only modestly when the first two indirect images are added.}
    \label{fig:mode_difference_photon_rings_extended}
\end{figure}

The band decomposition also allows us to ask whether the fast--slow discrepancy is driven by the indirect images or by the direct image. Figure~\ref{fig:mode_difference_photon_rings_extended} compares the standardized slow-light and fast-light curves after truncating the image at successive lensing orders. The norms show that the discrepancy is already set mainly by the direct image. Adding the first indirect image changes the integrated distances modestly, and adding the second indirect image changes them only slightly. 

\begin{figure}[]
    \centering
    \includegraphics[width=\linewidth]{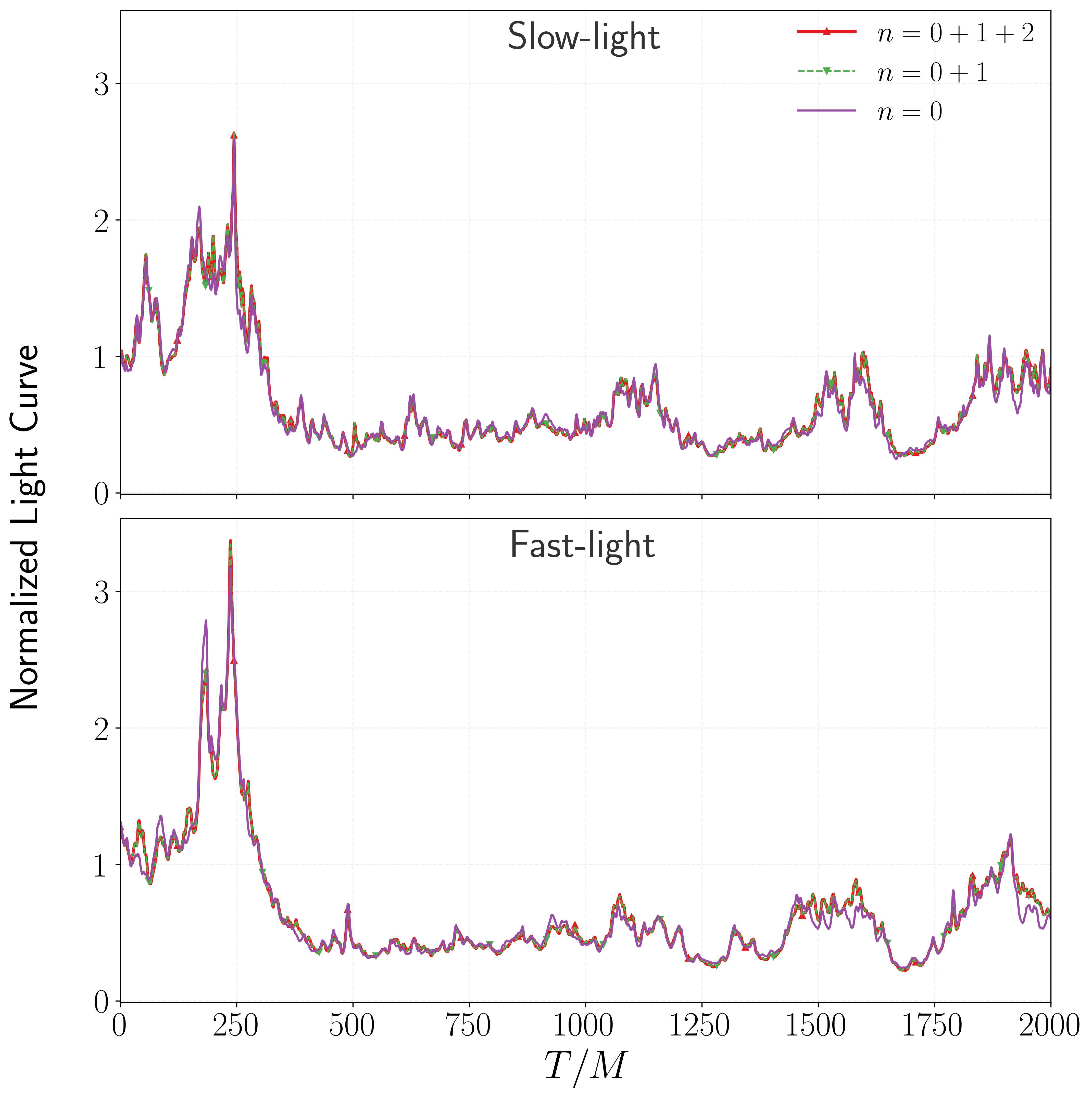}
    \caption{Mean-normalized light curves obtained by cumulative band addition. The upper panel shows slow light and the lower panel fast light. In both prescriptions, adding the first indirect image produces the largest correction to the direct-image light curve; the second indirect image is smaller for this configuration.}
    \label{fig:PR_Curve}
\end{figure}

Figure~\ref{fig:PR_Curve} shows how the light curves change under cumulative band addition. The transition from $n=0$ to $n\leq1$ produces the largest correction to the direct-image light curve, while the incremental $n=2$ contribution is subdominant for the unresolved flux in this configuration.

Higher-order images, however, can become more important for different optical depths, observing frequencies, source morphologies, or inclinations~\cite{Vincent:2022ThickDisks,DesireAstroModels}. The point is methodological: once the bandwise delay and flux structure are known, one can determine whether indirect images matter dynamically in a given black-hole movie.

\section{Brisk light}
\label{sec:brisk}

The delay histograms in Fig.~\ref{fig:Hist_17_log} suggest an intermediate prescription, which we call \emph{brisk light}: a band-delayed prescription that takes advantage of this behavior. Rather than collapsing the full image to a single source time, as in fast light, or retaining the full screen-dependent delay field, as in slow light, brisk light preserves the dominant temporal interval of each lensing band and clips only the low-density tails. It is therefore a bandwise, geometry-guided support reduction of slow light. 

\begin{figure}[]
    \centering
    \includegraphics[width=\linewidth]{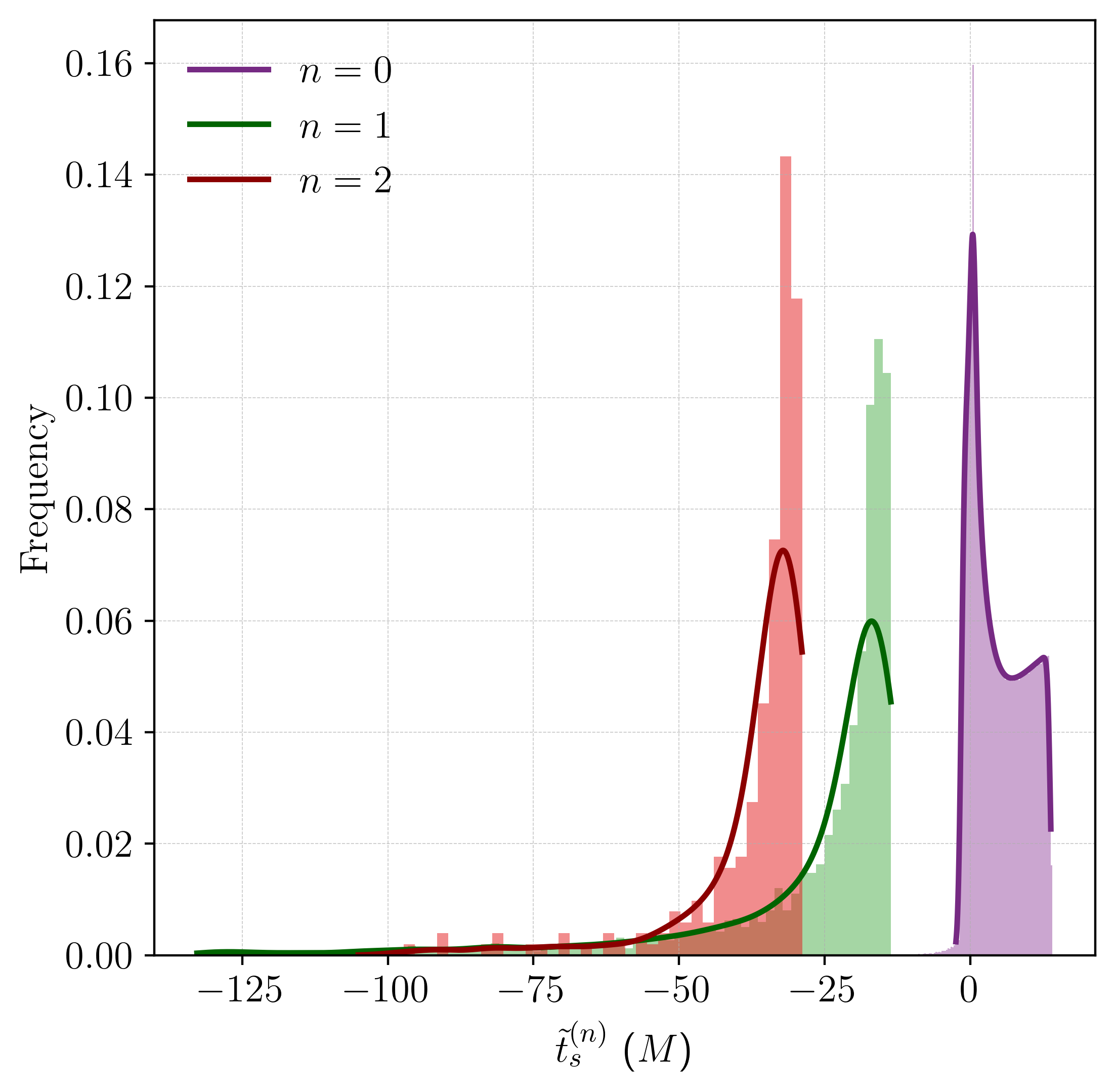}
    \caption{Normalized delay distributions for the first three lensing bands at $\theta_{\rm o}=17^\circ$, with Gaussian-kernel density estimates (KDE) overlaid. The KDEs are used to define the modal highest-density intervals used in the brisk-light prescription. The important feature is not only the separation between bands, but also the existence of a dense temporal core within each band.}
    \label{fig:Hist_17_kde}
\end{figure}

For the brisk-light construction, we use a smooth definition of the dominant delay support that preserves the delay map inside the modal core of each band while compressing the tails. Figure~\ref{fig:Hist_17_kde} shows the same distributions as Fig.~\ref{fig:Hist_17_log}, but with each histogram normalized as a probability density, so that the total area under each distribution is unity. A Gaussian kernel density estimate (KDE) is overlaid on each histogram.

For each band $n$, let $\hat f_n(t)$ be a Gaussian-KDE estimate of the probability density of sampled emission times, and let $\bar t_n$ denote its mode. For $0<p<1$, we define the modal highest-density interval (HDI) $\mathcal{T}_{n,p}$ as the connected interval containing the KDE mode $\bar t_n$ and enclosing probability mass $p$:
\begin{equation}
    \int_{\mathcal{T}_{n,p}}\hat f_n(t)\,dt=p.
    \label{eq:hdi_mass_condition}
\end{equation}

Operationally, $\mathcal{T}_{n,p}$ is obtained by lowering a density threshold around the modal peak until the connected interval containing $\bar t_n$ encloses mass $p$. For multimodal KDEs, the brisk-light construction selects the connected interval associated with the global mode. The limiting cases are implemented as $\mathcal{T}_{n,0}=\{\bar t_n\}$ and as $\mathcal{T}_{n,1}$ equal to the full sampled delay support of the band. Thus, $p$ controls the amount of delay probability retained around the modal time.

For a general source model, we take $\mathcal{T}_{n,p}=[t_-^{(n)},t_+^{(n)}]$ to be the modal highest-density interval on the source-time axis. Brisk light then clips the slow-light emission time to the nearest point in this interval. We denote this clipping map by $\mathcal{C}_{\mathcal{T}_{n,p}}$, so that
\begin{equation}
    \hat t_{\rm s}^{(n)}(t_{\rm o},\alpha,\beta)
    =\mathcal{C}_{\mathcal{T}_{n,p}}\!\left[t_{\rm s}^{(n)}(t_{\rm o},\alpha,\beta)\right].
    \label{eq:clipped_time_map}
\end{equation}
Here $\mathcal{C}_{\mathcal{T}_{n,p}}(t)=t$ for $t\in\mathcal{T}_{n,p}$ and $\mathcal{C}_{\mathcal{T}_{n,p}}(t)$ equals the nearer of the two endpoints when $t\notin\mathcal{T}_{n,p}$. The brisk-light image is
\begin{equation}
    \begin{aligned}
        I_{{\rm o},{\rm b}}^{(n)}(t_{\rm o},\alpha,\beta)
        &=g_n^3(\alpha,\beta)\,
        I_{\rm s}\!\left[\hat t_{\rm s}^{(n)}(t_{\rm o}),
        r_{\rm s}^{(n)},\phi_{\rm s}^{(n)}\right],
    \end{aligned}
    \label{eq:brisk_image_definition}
\end{equation}
and the total brisk image is obtained by summing over $n$ as in Eq.~\eqref{eq:total_observed_image}. The geodesics, redshift factors, and source positions are unchanged; only the temporal argument of the source intensity is compressed.

There is one practical point in this construction that is important for higher-order lensing bands. Formally, the source-time support of an indirect band can be very extended: the corresponding screen region contains rays whose equatorial emission points approach both the near-horizon region and very large disk radii. At finite image resolution, a small number of pixels can therefore return extremely large values of the renormalized emission time. These points are not representative of the temporal core of the image, but if included directly in the KDE they can make the estimated support of $\hat f_n(t)$ artificially broad and cause the HDI to be controlled by a handful of tail samples.

Consequently, we construct the KDE from a trimmed finite sample. After removing non-finite values, we retain the central fraction $q$ of the sampled emission times in each band and use only this subset to estimate $\hat f_n(t)$ and define $\mathcal{T}_{n,p}$. Equivalently, if $Q_n(u)$ denotes the $u$-th quantile of the finite bandwise time sample, we use
\begin{equation}
    Q_n\!\left(\frac{1-q}{2}\right)
    \leq
    t_{\rm s}^{(n)}
    \leq
    Q_n\!\left(\frac{1+q}{2}\right)
\end{equation}
when constructing the KDE. Unless otherwise stated, we take $q=0.995$. This value was sufficient in the high-spin and high-inclination cases considered here, where the long-time tails are most pronounced.

This trimming is used only to define the smooth delay distribution and its modal HDI; it does not remove pixels from the image. The original ray-traced pixels are still included in the brisk-light image, and emission times outside $\mathcal{T}_{n,p}$ are treated as tail points and clipped to the nearest HDI boundary. The probability mass $p$ should therefore be understood as the mass enclosed with respect to the trimmed KDE. In the ideal untrimmed definition, $p=1$ recovers the finite-sample slow-light support. In the trimmed numerical implementation, the exact slow-light result is obtained by bypassing the clipping map and using the full ray-traced time map directly.

Brisk light is not a bandwise quantization of the entire image. For $p>0$, all emission times already inside the modal interval are kept exactly. Only the tails are clipped to the nearest boundary. In this sense, brisk light occupies an intermediate position in temporal information content: fast light retains one global source time, brisk light retains a bandwise modal support, and slow light retains the full screen-dependent time map. This ordering should be understood as a hierarchy of temporal support, not as a literal set inclusion of images.

In the ideal untrimmed definition, when $p=1$, the modal interval covers the full support and $\hat t_{\rm s}^{(n)}=t_{\rm s}^{(n)}$, so brisk light reduces to the discrete slow-light calculation. In the trimmed numerical implementation used above, the exact slow-light result is obtained by bypassing the clipping map and using the full ray-traced time map directly. When $p=0$, every pixel in a given band is evaluated at the modal time $\bar t_n$. This is still not fast light unless all band modes coincide. The modal brisk image uses one representative source time per retained lensing band, ordered by the geometric delays $\bar t_0>\bar t_1>\bar t_2>\cdots$. Even in this most compressed limit, brisk light preserves the leading temporal offset between direct and indirect emission, whereas fast light synchronizes the entire screen.

The computational gain is therefore different from that of fast light. The ray tracing still visits the same screen points. The saving comes from reducing the temporal support needed by the source interpolation or snapshot-loading stage. For a stored source movie, $p$ specifies how much of the bandwise delay support must remain available; larger $p$ approaches slow light, smaller $p$ approaches a small set of band-selected times.

\subsection{Comparisons to the slow and fast prescriptions}

We compare brisk light to slow light using the same $L^1$ and $L^2$ distances defined in Eqs.~\eqref{eq:l1_def} and~\eqref{eq:l2_def}, with slow light as the reference. Since $p$ is a probability-mass parameter, it need not correspond to the same number of retained source times at different inclinations. A broader, flatter delay distribution generally requires a wider interval, and therefore a larger effective temporal support, to enclose the same $p$. For this reason, $p$ should be interpreted as a controlled accuracy parameter rather than a universal fixed-cost parameter.

\begin{figure}[]
    \centering
    \includegraphics[width=\linewidth]{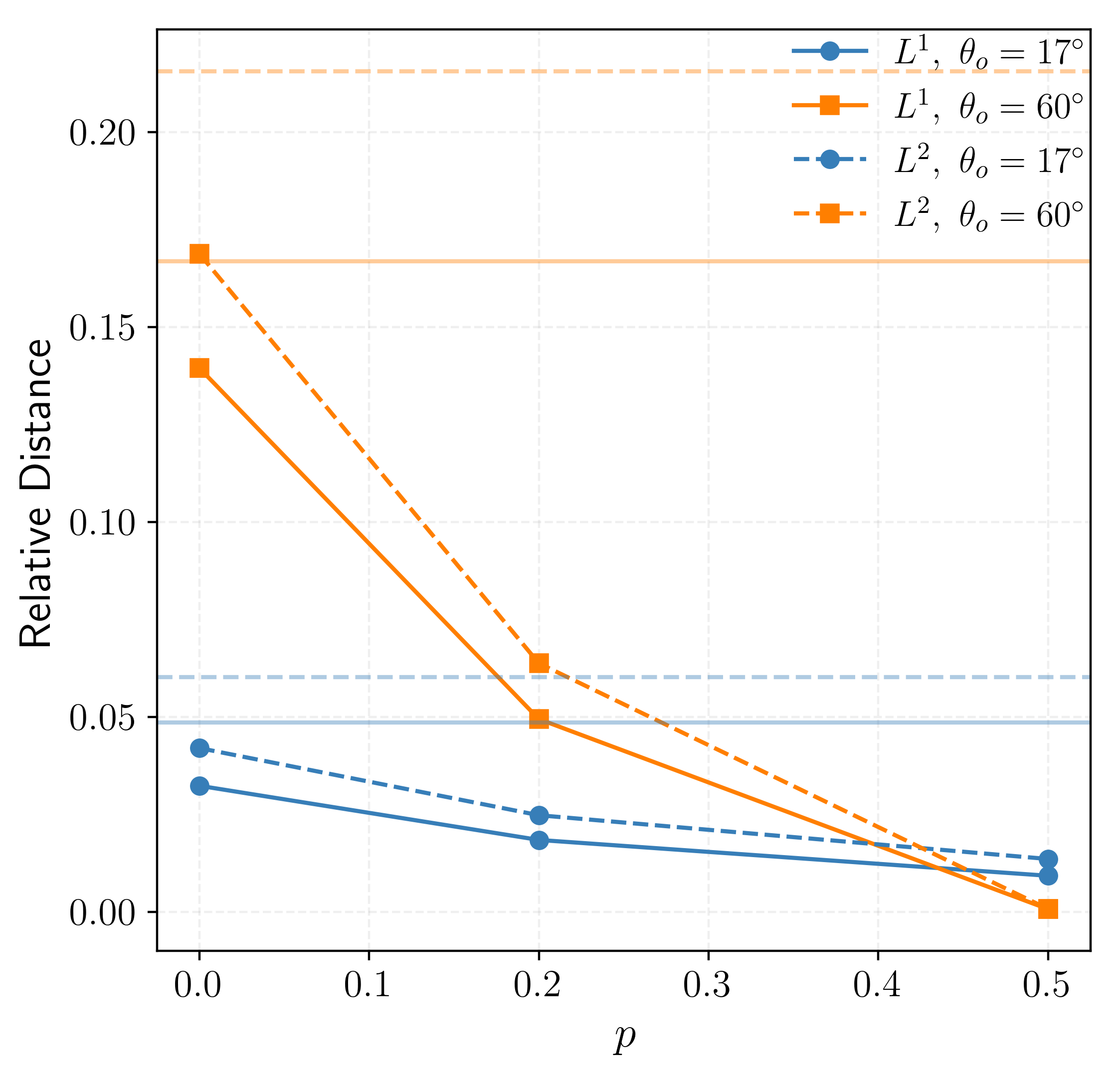}
    \caption{Relative $L^1$ and $L^2$ distances between standardized brisk-light and slow-light curves as functions of the enclosed probability mass $p$. Dashed horizontal lines show the corresponding fast-light distances. The modal brisk limit ($p=0$) already improves on fast light because it retains one dominant time per lensing band rather than one global time for the full image. The horizontal axis is therefore a probability-mass, or accuracy, parameter rather than a fixed computational-cost parameter.}
    \label{fig:ML1L2_p}
\end{figure}

For the standardized integrated light curves shown in Fig.~\ref{fig:ML1L2_p}, brisk light is closer to slow light than fast light even at $p=0$. This result isolates the most important difference between the two approximations. Fast light uses a single source snapshot for all bands. Modal brisk light uses a small number of band-resolved snapshots, displaced relative to each other by the Kerr delay structure. The improvement therefore comes from retaining the leading temporal ordering of the image, not from changing the geodesics or the source model.

\begin{figure}[]
    \centering
    \includegraphics[width=\linewidth]{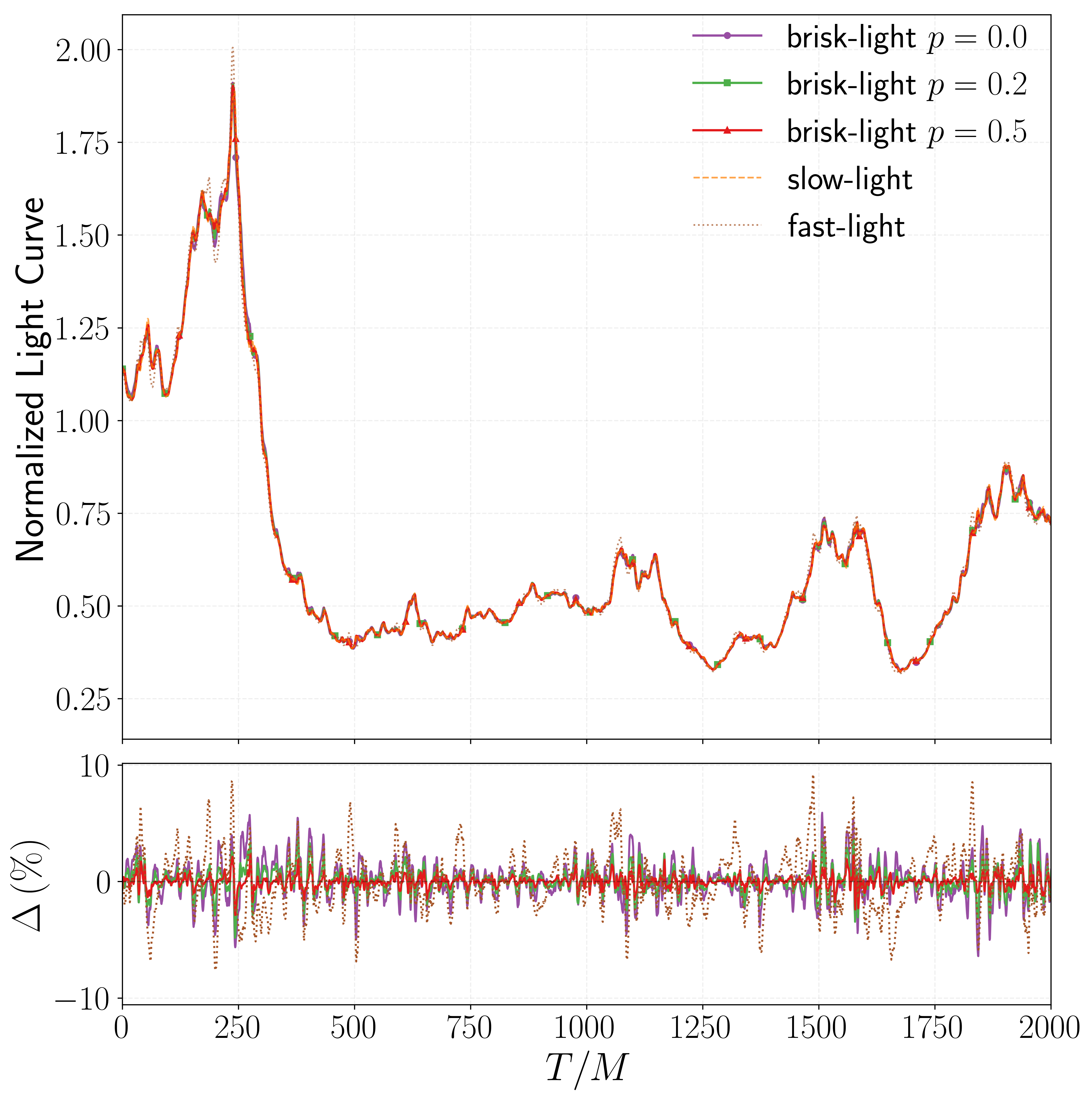}
    \caption{Brisk-light convergence at $\theta_{\rm o}=17^\circ$. Top: mean-normalized light curves for slow light (dashed orange line), fast light (dotted brown line), and brisk light (solid lines) at several values of $p$. Bottom: pointwise relative difference between each brisk-light curve and the slow-light reference. Because the low-inclination delay distribution is already compact, fast light is accurate and the remaining brisk-light corrections are small.}
    \label{fig:LC_Brisk_17}
\end{figure}

\begin{figure}[]
    \centering
    \includegraphics[width=\linewidth]{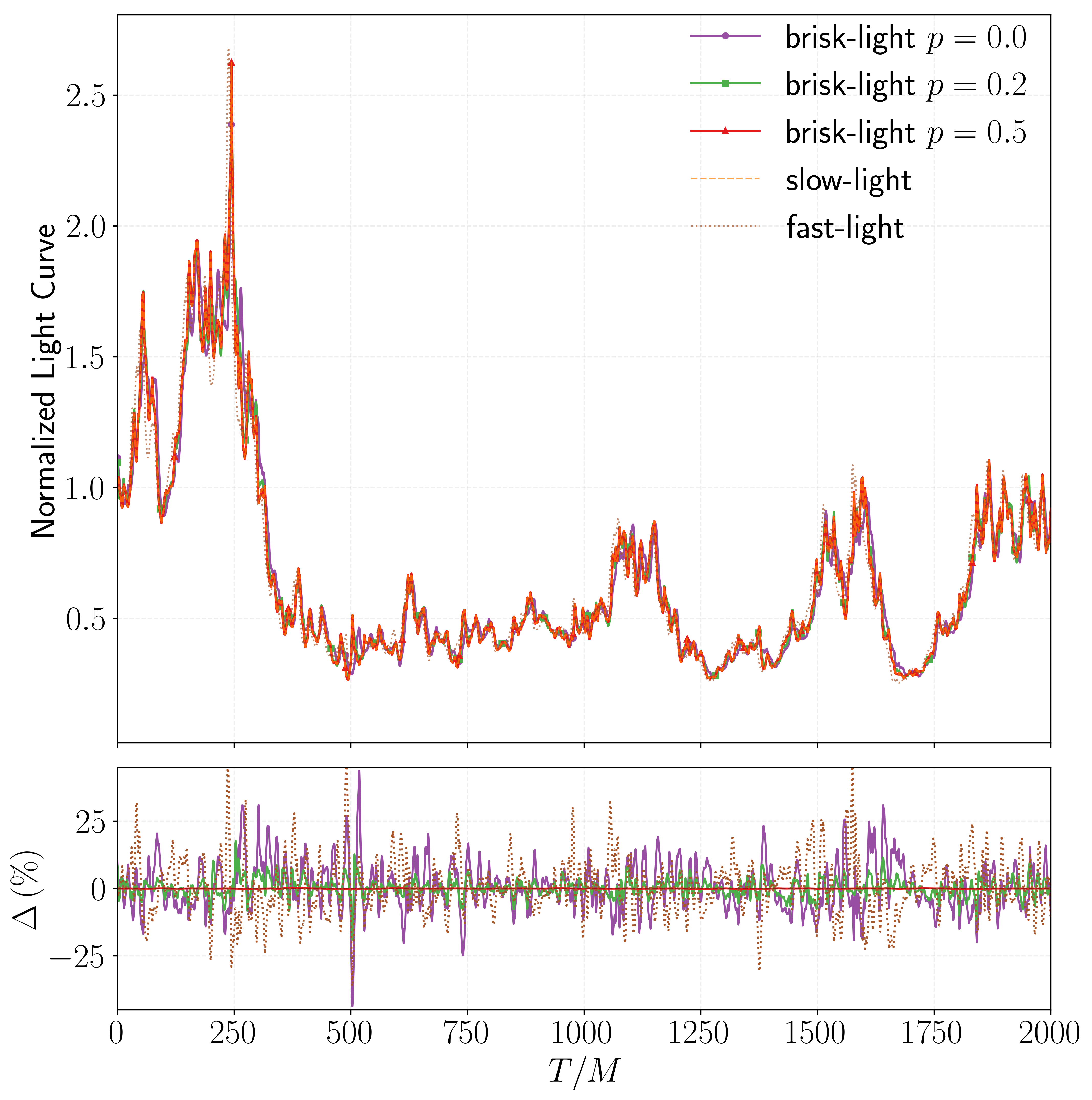}
    \caption{Same as Fig.~\ref{fig:LC_Brisk_17}, but for $\theta_{\rm o}=60^\circ$. The modal limit is less accurate than at low inclination, as expected from the broader delay distribution, but increasing $p$ rapidly suppresses the residuals because the HDI recovers a much larger fraction of the relevant delay map.}
    \label{fig:LC_Brisk_60}
\end{figure}

Figures~\ref{fig:ML1L2_p},~\ref{fig:LC_Brisk_17}, and~\ref{fig:LC_Brisk_60} show two complementary trends. At low inclination (Fig.~\ref{fig:LC_Brisk_17}), fast light is already accurate because the direct-image delay distribution is compact. Consequently, increasing $p$ has less room to improve the light curve. At high inclination (Fig.~\ref{fig:LC_Brisk_60}), fast light performs worse because the delay field is broader and more structured; each increase in $p$ then adds a substantial amount of previously clipped delay support. Brisk light therefore improves most rapidly in the regime where fast light loses the most temporal information.

This faster convergence should be interpreted with one practical caveat. The same value of $p$ does not necessarily imply the same cost. A broad high-inclination HDI can contain more discrete source snapshots than a narrow low-inclination HDI. Thus, the high-inclination curves improve quickly both because the additional delay support is physically important and because a fixed probability mass may correspond to a larger effective number of retained source times. This is why a complete computational benchmark should report both $p$ and the effective retained temporal support.

\subsection{Image-domain comparison}

\begin{figure*}[]
    \centering
    \includegraphics[width=\linewidth]{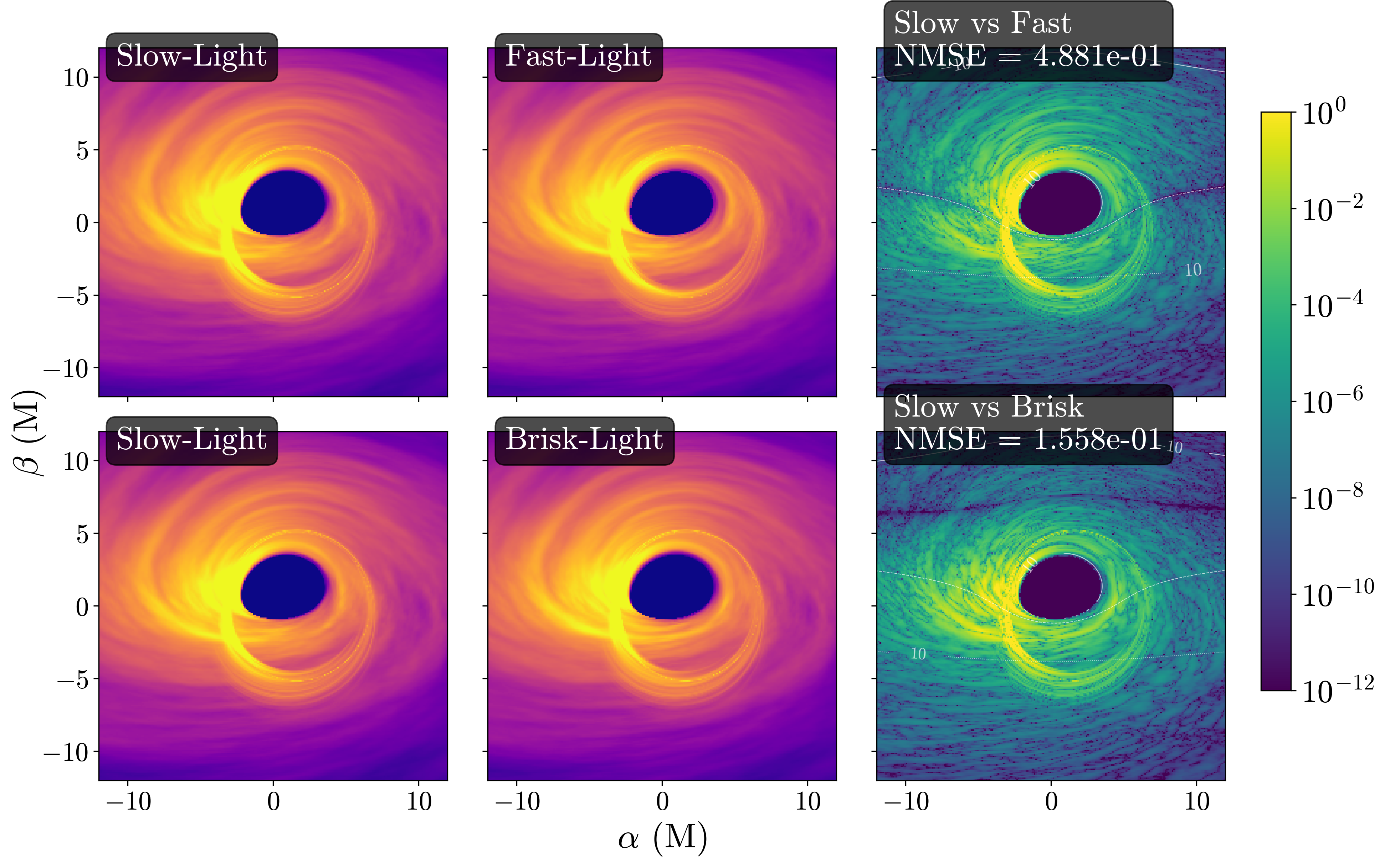}
    \caption{Image-domain comparison at $\theta_{\rm o}=60^\circ$ in the modal brisk limit $p=0$. The top row compares slow and fast light, while the bottom row compares slow and brisk light. The first two columns in each row show the corresponding intensity maps on the same screen, while the third column shows the squared pixelwise difference $\delta(i,j)$ on a logarithmic color scale; the NMSE is reported in each residual panel. White contours mark isochrones of the renormalized Kerr emission-time map. Dark ridges in the residual panels occur where the two prescriptions evaluate the source at the same, or nearly the same, emission time.}
    \label{fig:BHPComp_p0i60}
\end{figure*}

\begin{figure*}[]
    \centering
    \includegraphics[width=\linewidth]{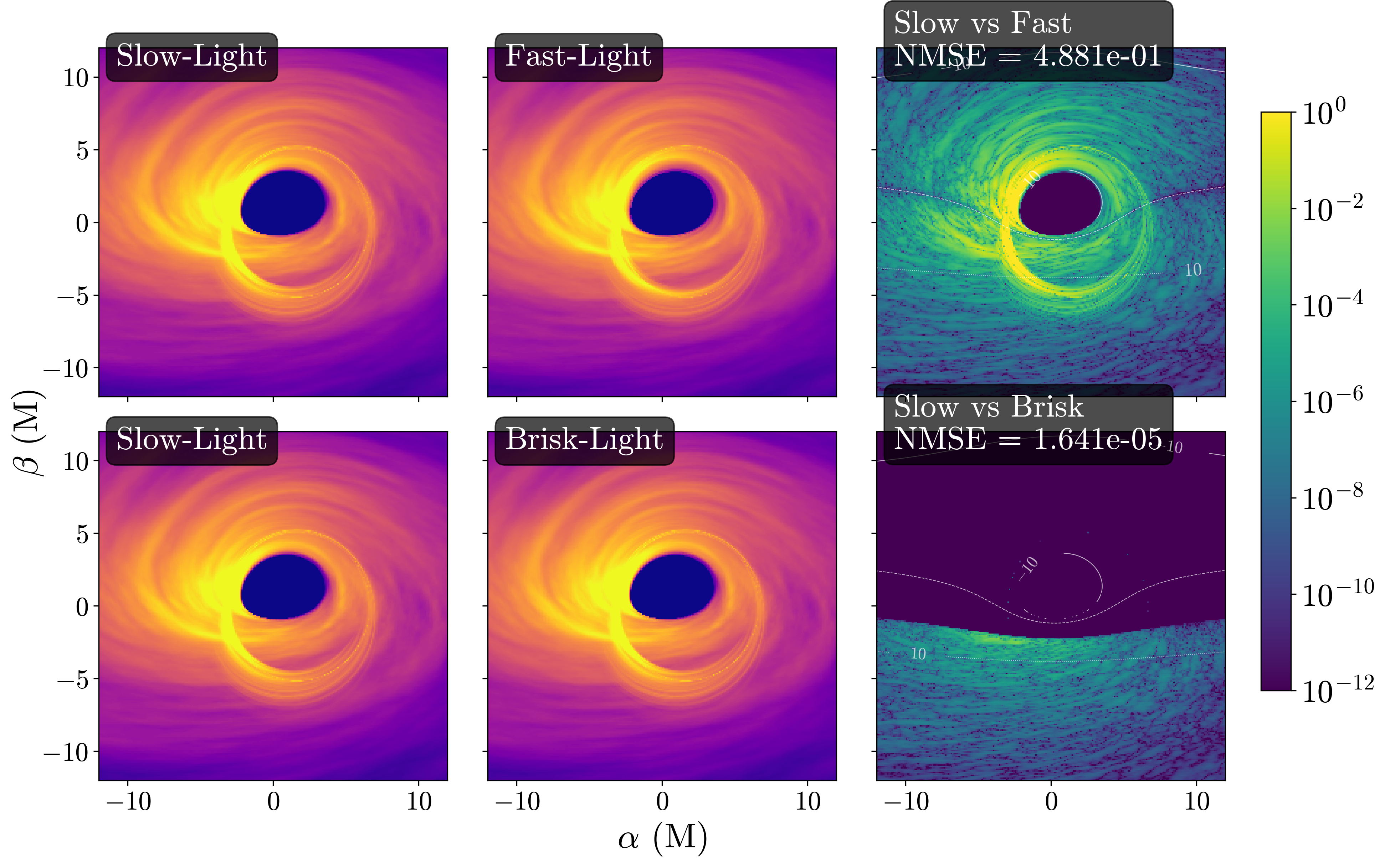}
    \caption{Same image-domain comparison as in Fig.~\ref{fig:BHPComp_p0i60}, but for $p=0.5$. Widening the HDI allows brisk light to retain a much larger fraction of the slow-light temporal map, so the slow--brisk residual, shown in the bottom row, is suppressed by several orders of magnitude relative to the slow--fast residual. The remaining slow--brisk differences are localized mainly in the direct image. At the displayed resolution, the part of the photon-ring region not covered by the retained HDI is sufficiently small that only a handful of pixels remain visible on the logarithmic residual scale.}
    \label{fig:BHPComp_p05i60}
\end{figure*}

The image-domain residuals show where the prescriptions differ on the screen. Figures~\ref{fig:BHPComp_p0i60} and~\ref{fig:BHPComp_p05i60} compare snapshots at $\theta_{\rm o}=60^\circ$ for $p=0$ and $p=0.5$, respectively. For two images $I_A$ and $I_B$ on the same pixel grid, we report the normalized mean-squared error
\begin{equation}
    \mathrm{NMSE}(I_A,I_B)=
    \frac{\displaystyle\sum_{i,j}\left[I_A(i,j)-I_B(i,j)\right]^2}
    {\displaystyle\sum_{i,j}I_A(i,j)^2},
    \label{eq:nmse}
\end{equation}
and display the squared difference map
\begin{equation}
    \delta(i,j)=\left[I_A(i,j)-I_B(i,j)\right]^2.
    \label{eq:pixel_difference}
\end{equation}
The NMSE is useful as a compact scalar but should not be read as a complete morphology metric: it is flux-weighted, sensitive to normalization, and sensitive to small spatial shifts of bright structures~\cite{RoelofsImaging}. As shown in Figs.~\ref{fig:LC_Brisk_17} and~\ref{fig:LC_Brisk_60}, brisk light is not guaranteed to outperform fast light for every source realization, cadence, or diagnostic. The $L^1$ and $L^2$ distances are global light-curve metrics, and their values can be affected by realization-dependent cancellations, phase alignments, and the particular temporal structure of the emission. Nevertheless, the overall trend is that brisk light provides a more faithful approximation to slow light: by retaining the dominant bandwise delay structure, it generally reduces the integrated light-curve mismatch and suppresses the largest pointwise residuals relative to fast light.

Since the geodesics are identical in all prescriptions, the residual structure is produced entirely by differences in the source time used at each pixel. In a fast--slow comparison, the residual is small along the isochrone $t_{\rm s}^{(n)}(\alpha,\beta)=t_{\rm f}$, because both prescriptions sample the same source state there. In a brisk--slow comparison, the low-residual region is the part of the screen whose emission times lie inside, or are close to, the modal interval $\mathcal{T}_{n,p}$. Increasing $p$ widens that region, which is exactly what is seen when comparing Figs.~\ref{fig:BHPComp_p0i60} and~\ref{fig:BHPComp_p05i60}.

\begin{figure}[]
    \centering
    \includegraphics[width=\linewidth]{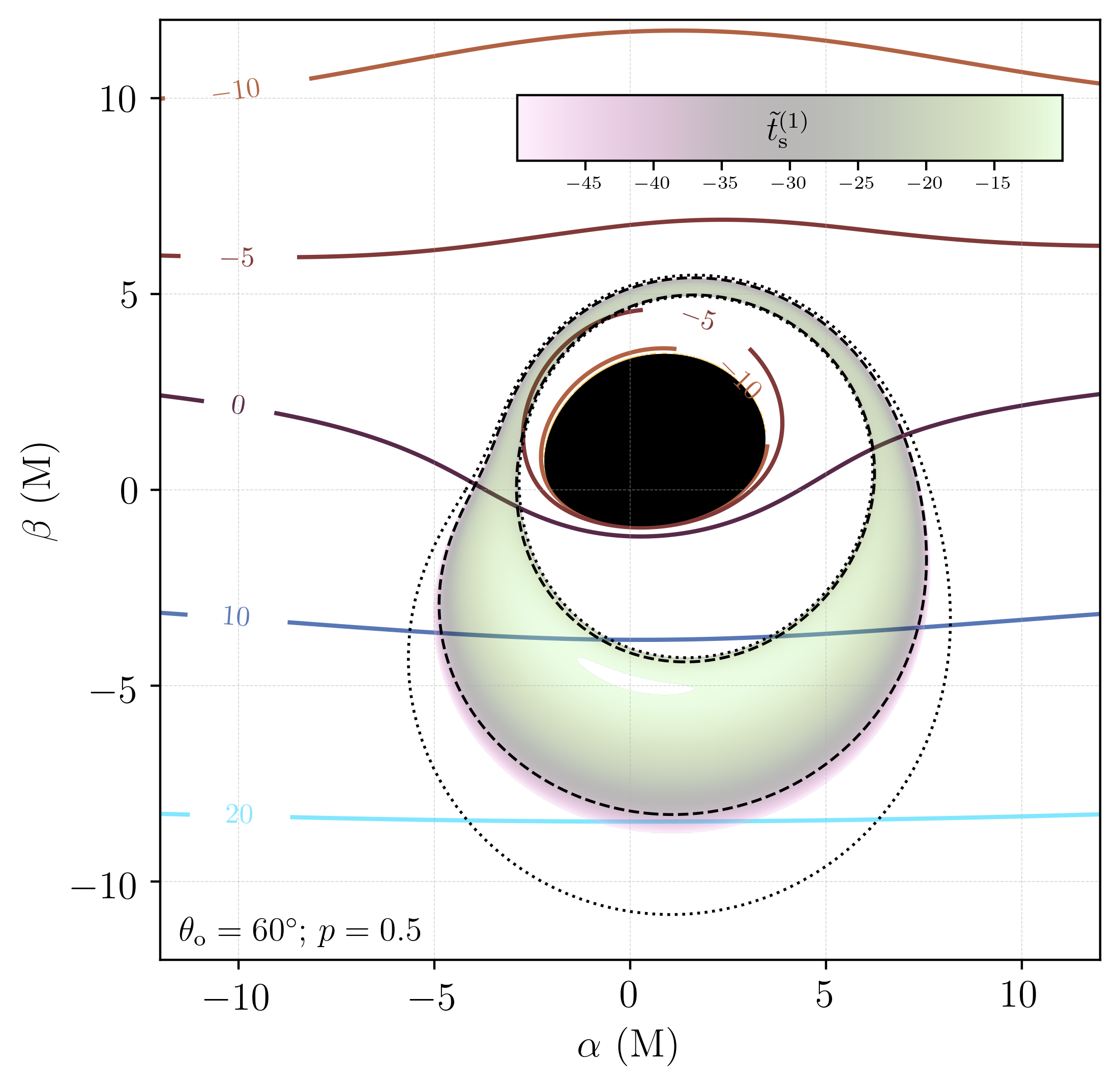}
    \caption{Isochrones and brisk-light temporal support at $\theta_{\rm o}=60^\circ$ for $p=0.5$. Solid contours show the renormalized direct-image emission time $\tilde t_{\rm s}^{(0)}$, measured after subtracting the observer-radius contribution for $r_{\rm o}=10^4M$. The shaded region shows the subset of the first indirect lensing band whose renormalized emission times $\tilde t_{\rm s}^{(1)}$ lie inside the modal highest-density interval $\mathcal{T}_{1,0.5}$; the color scale gives the corresponding value of $\tilde t_{\rm s}^{(1)}$. Pixels in this shaded region are sampled by brisk light at their slow-light emission time, while pixels outside the retained interval are clipped to the nearest HDI boundary. The black dotted curves mark the boundaries of the $n=1$ lensing band, the black dashed curves mark the screen footprint of the finite equatorial emitting region, and the black filled region denotes the image of the horizon. The large overlap between the retained $p=0.5$ support and the first indirect image explains why the slow--brisk residual is strongly reduced relative to the slow--fast residual in the image-domain comparison.}
    \label{fig:IsoBriskSupport}
\end{figure}

Figure~\ref{fig:IsoBriskSupport} makes the interpretation of the brisk prescription explicit for the cases shown in Figs.~\ref{fig:BHPComp_p0i60} and~\ref{fig:BHPComp_p05i60}. The solid curves show the direct-image isochrones $\tilde t_{\rm s}^{(0)}={\rm const.}$, while the shaded region shows the portion of the first indirect band for which the emission time $\tilde t_{\rm s}^{(1)}$ is retained by brisk light at $p=0.5$. In this example, the retained interval spans a broad range of first-band source times, roughly from $\tilde t_{\rm s}^{(1)}\simeq -50M$ to $-10M$, so brisk light is not evaluating the first indirect image at a single time; it is preserving a substantial part of the slow-light time map. This is why the slow--brisk residual in the image-domain comparison is reduced by several orders of magnitude relative to the slow--fast residual once $p=0.5$.

As Fig.~\ref{fig:IsoBriskSupport} shows, in a fast--slow comparison, the prescriptions agree (darkest regions in Figs.~\ref{fig:BHPComp_p0i60} and~\ref{fig:BHPComp_p05i60}) along the isochrone corresponding to the chosen fast-light reference time, here $\tilde t_{\rm s}=0$. In the modal brisk limit, however, the relevant time is the mode $\bar t_n$ of the bandwise distribution, not the arbitrary zero of the renormalized time coordinate. The dark ridge in a brisk--slow residual map can therefore appear at a different screen location from the fast--slow ridge.

These image comparisons also clarify why brisk light can improve rapidly with $p$ at higher inclination. The complementary low-inclination behavior is consistent with the structure of the isochrones: near face-on, closed isochrones cover a larger fraction of the direct-image region, including the area around the inner shadow~\cite{Chael:2021InnerShadow}. Compressing the $n=0$ delay field can therefore create coherent residuals over a broad screen area. At higher inclination, the delay families are more extended and separated; once the HDI is widened, brisk light recovers the dominant families more efficiently.

\section{Discussion}
\label{sec:discussion}

In this work, we have shown that the relevance of slow-light propagation, the relativistically faithful prescription for black hole imaging, is controlled by the competition between source variability and the geodesic delay structure. Fast-light is accurate when the source remains correlated across the delay support of the image, because replacing the screen-dependent emission-time map with a single source time removes little physical information. It deteriorates once the relevant lensing-band delay widths become comparable to, or larger than, the timescale over which the source changes appreciably. This criterion should be evaluated using the delay support that contributes appreciably to the observable of interest, which may be pixel-weighted, transfer-weighted, or flux-weighted depending on the application. This explains why near face-on cases are comparatively accurate, while higher-inclination cases show substantially larger fast--slow discrepancies.

The unresolved fast--slow light-curve difference is driven mainly by the direct image: adding the first two indirect images changes the total-flux metrics only modestly. This should not be interpreted as evidence that higher-order images are observationally unimportant, nor as evidence that slow-light effects are uniformly irrelevant for parameter inference, because total flux is not the relevant figure of merit for much of photon-ring science. Indirect images can contribute little to the unresolved light curve while still carrying the information most directly tied to strong-field null geodesics, including their relative delays, phase ordering, and correlations across lensing order. In other words, although the photon-ring contribution can be a mild correction to the total flux, a temporally faithful treatment of this contribution is essential for observables that depend on delays, correlations, or light echoes~\cite{Johnson:2019UniversalInterferometric,Hadar:2020PhotonAutocorr,Wong:2020ziu,Wong:2024LightEchoesVLBI,Bezdekova:2025CorrelationsMovies}.

We emphasize, however, that the type of light propagation need not affect all observables or inferences in the same way. In the fast-, brisk-, and slow-light calculations considered here, the null geodesics, source positions, redshift factors, and critical curve are identical; what changes is the retarded source time at which the emissivity is sampled. In particular, a fast-light calculation retains the higher-order geodesics themselves, and assigning them a common source time does not erase their geometric information. Consequently, observables dominated by time-averaged image geometry, such as the angular scale of the emission ring used in current mass-scale constraints for M87* and Sgr A*~\cite{EventHorizonTelescope:2019ggy,EventHorizonTelescope:2022wkp}, are expected to be comparatively insensitive to the propagation prescription alone. Similarly, spin constraints based primarily on quasi-static image geometry or time-averaged morphology, including proposed measurements with future high-resolution or space-based VLBI capabilities~\cite{Keeble:2025gbj,Farah:2025kpb,Astrometry2026}, should be less sensitive to the light-propagation prescription than spin constraints based on time-domain correlations, glimmer, or echo delays~\cite{Hadar:2020PhotonAutocorr,Wong:2020ziu,Wong:2024LightEchoesVLBI,Zhang:2025vyx}. Other image information that is not purely geometric, such as constraints on the magnetic-field properties of the emitting plasma, also depends on astrophysical modeling, and consequently the impact of the light-propagation prescription on such inferences must be assessed case by case. For example, for evolving relativistic plasma, the light-propagation prescription can modify image morphology and polarization-related structures~\cite{EventHorizonTelescope:2021srq,Tsunetoe:2026SlowLightM87}. Thus, a temporally faithful treatment can also matter beyond delay- and correlation-based photon-ring observables. A quantitative assessment of the impact of the light-propagation prescription on parameter estimation is left for future work.

To reduce the computational cost and improve the efficiency of observables and inference schemes that rely on the image's temporal information, we have proposed brisk light: a practical reduced-delay prescription for computing black-hole movies. It uses the ray-traced delay distributions to identify the dominant temporal support of each lensing band, then evaluates the same source movie with a clipped version of the slow-light time map. The prescription therefore leaves the geodesics, the source positions, and the redshift factors unchanged. Its only approximation is temporal: it clips the full slow-light delay map to a bandwise modal interval while leaving times inside that interval unchanged. Because a single global source snapshot erases the relative timing between lensing bands, even the modal brisk limit ($p=0$) can outperform fast light under this prescription.

At low inclination, the delay distribution is narrow and fast light is already close to slow light, so increasing the retained probability mass $p$ produces only modest additional improvement. At higher inclination, the delay distribution is broader and fast light loses more information; increasing the retained probability mass ($p$) then restores a larger fraction of the physically relevant delay support.

Under the present definition, a fixed value of $p$ is not a universal cost parameter. The source-side cost is controlled instead by the retained delay width $W_n(p)$, or equivalently by the number of source snapshots that must be loaded, stored, or interpolated over. Since $W_n(p)$ depends on the delay distribution, the speedup depends on the spin, inclination, source geometry, fluid cadence, image cadence, and grid resolution. Nevertheless, reducing $p$ reduces the temporal working set whenever the source-access or interpolation stage is a non-negligible part of the calculation. For example, for a black hole with $a/M=0.94$ observed at $\theta_{\rm o}=60^\circ$, brisk light with $p=0.99$ covers a delay range of approximately $80\,M$ in the $n=0$ band, whereas $p=0.5$ covers only about $35\,M$. For a source cadence $\Delta T\simeq0.61M$, this corresponds to roughly $130$ versus $60$ source times. Thus, in this example, reducing $p$ from $0.99$ to $0.5$ lowers the source-stage temporal support by about a factor of two, although the total wall-clock speedup depends on the relative cost of ray tracing, source loading, interpolation, radiative-transfer evaluation, and memory movement.

Although our results are restricted to equatorial emission, this restriction cleanly isolates the screen-dependent delay structure of the \texttt{AART} equatorial transfer map and also limits the range of possible slow-light effects. In a fully three-dimensional emitting flow, photons can sample a much richer set of spacetime paths before reaching the observer (see, e.g., Fig.~3 of Ref.~\cite{Wong:2024LightEchoesVLBI}, where isochrone contours are shown together with the corresponding GRMHD emission), and the emitting plasma itself can evolve relativistically along those paths. Related slow-light or finite-light-crossing-time effects have been found in particle-in-cell jet radiative transfer, shearing-spot models of M87 variability, and models the jet-launching region~\cite{MacDonald:2021ElectronsToJanskys,Jeter:2020M87Variability,SaizPerez:2025NGC1052,Tsunetoe:2026SlowLightM87}. Even when the emissivity remains concentrated near preferred regions of the flow (e.g., the equatorial plane as shown in Ref.~\cite{Astrometry2026}), the delay contours and evolving plasma sample a richer set of propagation times than an exactly equatorial model permits. We therefore expect the differences between fast, brisk, and slow light to become more pronounced when appreciable emission originates away from the equatorial plane or from relativistically moving jet material.

More broadly, the geometric-first viewpoint followed in this work is not tied to stochastic sources, to analytic ray tracing, or even to the particular equatorial lensing-band convention used in this work. The only required geometric input is a set of geodesics computed a priori from the observer screen, either analytically or numerically, together with a rule for grouping the multiple contributions to the observed image. In the present calculation this grouping is the \texttt{AART} lensing-band decomposition, based on equatorial-plane crossing order. Other applications may instead group rays by time ordering, by crossings of the equatorial plane or the $z=0$ plane, by polar or $\theta$ turning points, or by another decomposition adapted to the observable of interest~\cite{LensingBandsGeometric,Wong:2024LightEchoesVLBI}. The labels and detailed delay distributions will change with this choice, but the brisk-light construction does not: for each contribution, one computes the associated emission-time distribution, identifies its dominant temporal support, and compresses only the temporal argument of the source interpolation while leaving the geodesics, transfer factors, and source positions unchanged.

Beyond accreting black-hole systems, the present delay-compression viewpoint may also be useful in related time-domain lensing problems. For instance, in microlensing of fast transients, multiple micro-images can be unresolved angularly but separated in arrival time~\cite{Lewis:2020asm}. In that setting, the analogue of brisk light would not be a photon-ring-band prescription, but a compressed representation of the microlensing delay kernel: one would replace the full set of micro-image delays and magnifications by a smaller set of geometrically defined, magnification-weighted delay groups, while preserving the arrival-time ordering relevant for such systems. Similarly, an analogous construction based on \emph{timelike} geodesics, for which the delay map depends on the particle energy or asymptotic velocity~\cite{Frost:2023enn,Igata:2026hzb}, would be another interesting future study. Unlike in the null case considered here, the trajectory map itself would then depend on the messenger energy, and therefore a brisk-like approximation could compress not only the arrival-time distribution, but also the relevant family of timelike trajectories. 

Future work will be devoted to applying this definition-independent viewpoint in settings where the geodesic grouping and the emission model are both more complex. In particular, we leave to future work a systematic assessment of brisk light in fully numerical GRMHD movies and in observables beyond the integrated flux, especially visibility-domain and correlation-based probes aimed at detecting indirect emission. Such observables are especially sensitive to the temporal ordering imposed by lensing, and therefore provide a natural setting in which the distinction between fast, brisk, and slow light may become most pronounced.

\acknowledgments

We thank C.~Gammie, S.~Hadar, L.~Keeble, and P.~Motta for useful comments and discussions. This work used Delta at NCSA through allocation PHY250091 from the Advanced Cyberinfrastructure Coordination Ecosystem: Services \& Support (ACCESS) program, which is supported by U.S. National Science Foundation grants \#2138259, \#2138286, \#2138307, \#2137603, and \#2138296.

\appendix

\section{Convergence of the source and image discretization}
\label{app:convergence}

Before interpreting propagation effects, we verify that the comparison is not dominated by the numerical resolution of either the ray-traced image or the source movie. We varied the spatial ray-tracing resolution $\Delta x$, the source cadence $\Delta T$, and the ray-tracing cadence $\Delta t$ independently. Here $\Delta T$ is the spacing between stored source snapshots, and $\Delta t$ is the spacing between observer frames.

\begin{figure}[h!]
    \centering
    \includegraphics[width=\linewidth]{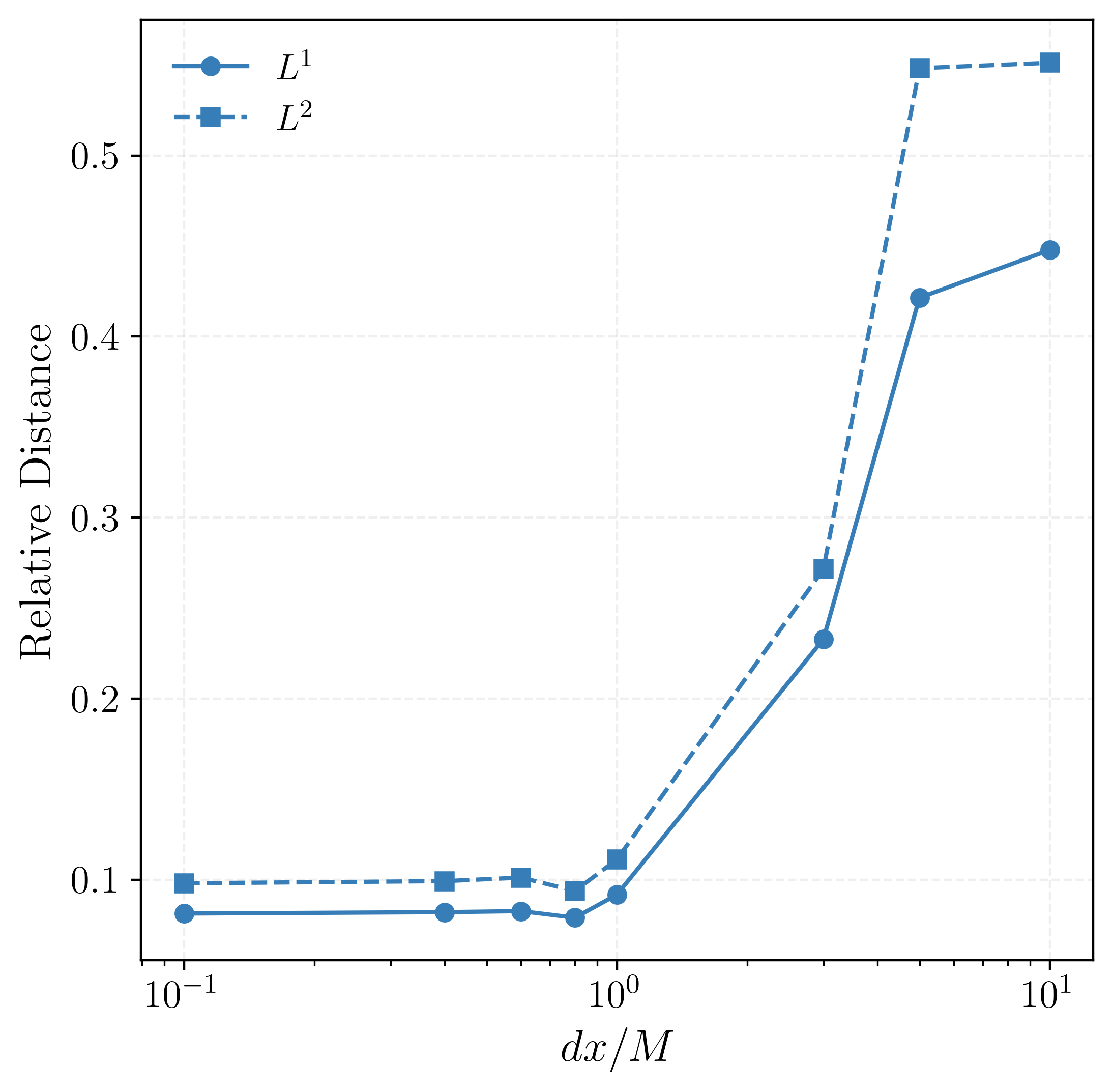}
    \caption{Relative $L^1$ and $L^2$ distances between standardized slow-light and fast-light light curves as functions of the screen resolution $\Delta x$. The plateau for $\Delta x\lesssim1M$ indicates that the reported propagation differences are not set by the screen discretization.}
    \label{fig:ErrL1L2_dx}
\end{figure}

\begin{figure}[]
    \centering
    \includegraphics[width=\linewidth]{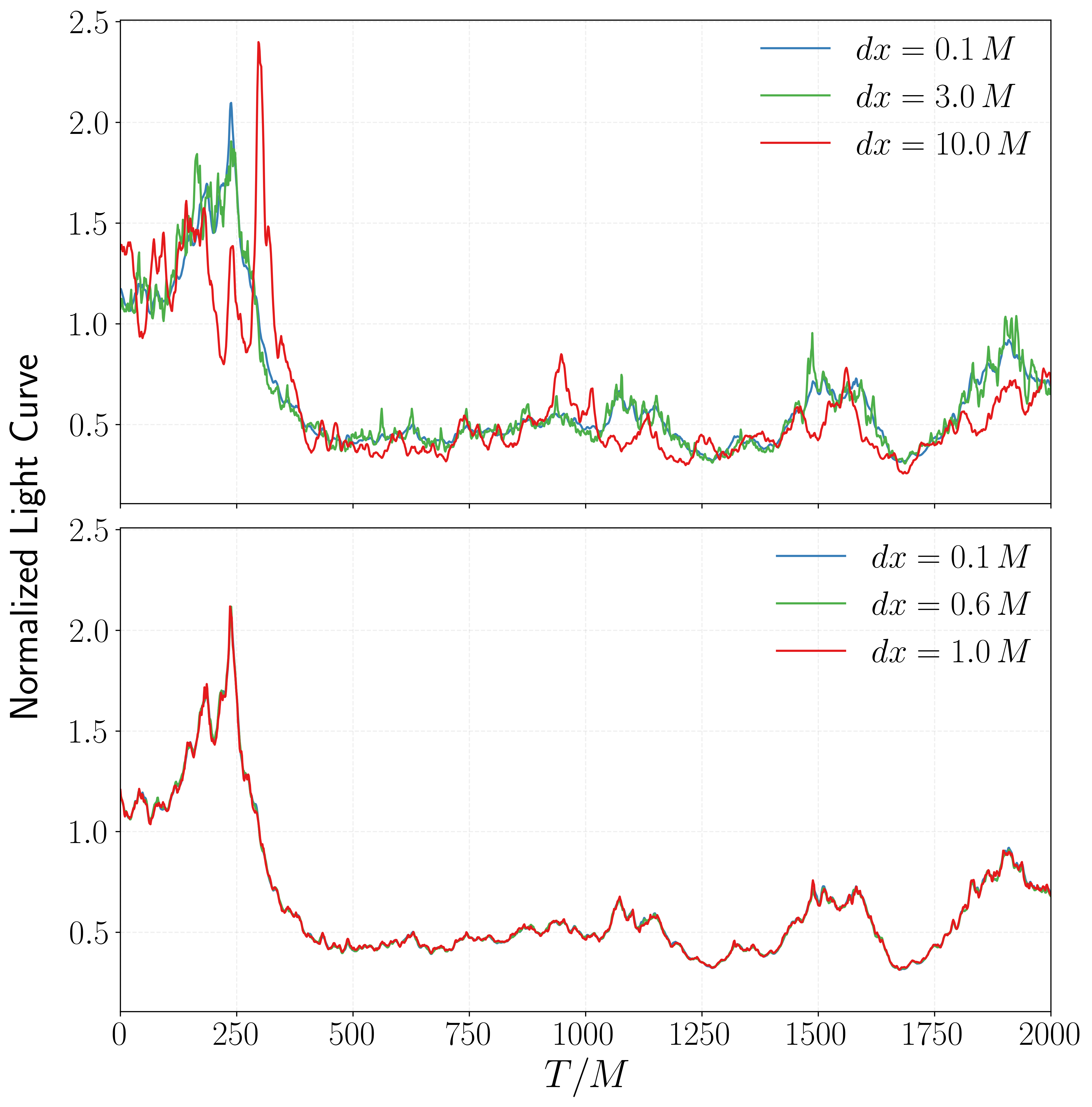}
    \caption{Mean-normalized fast-light curves at several screen resolutions. The upper panel includes coarse grids that visibly alter the light curve; the lower panel zooms into the converged range, where the curves are nearly indistinguishable.}
    \label{fig:FL_dx}
\end{figure}

\begin{figure}[]
    \centering
    \includegraphics[width=\linewidth]{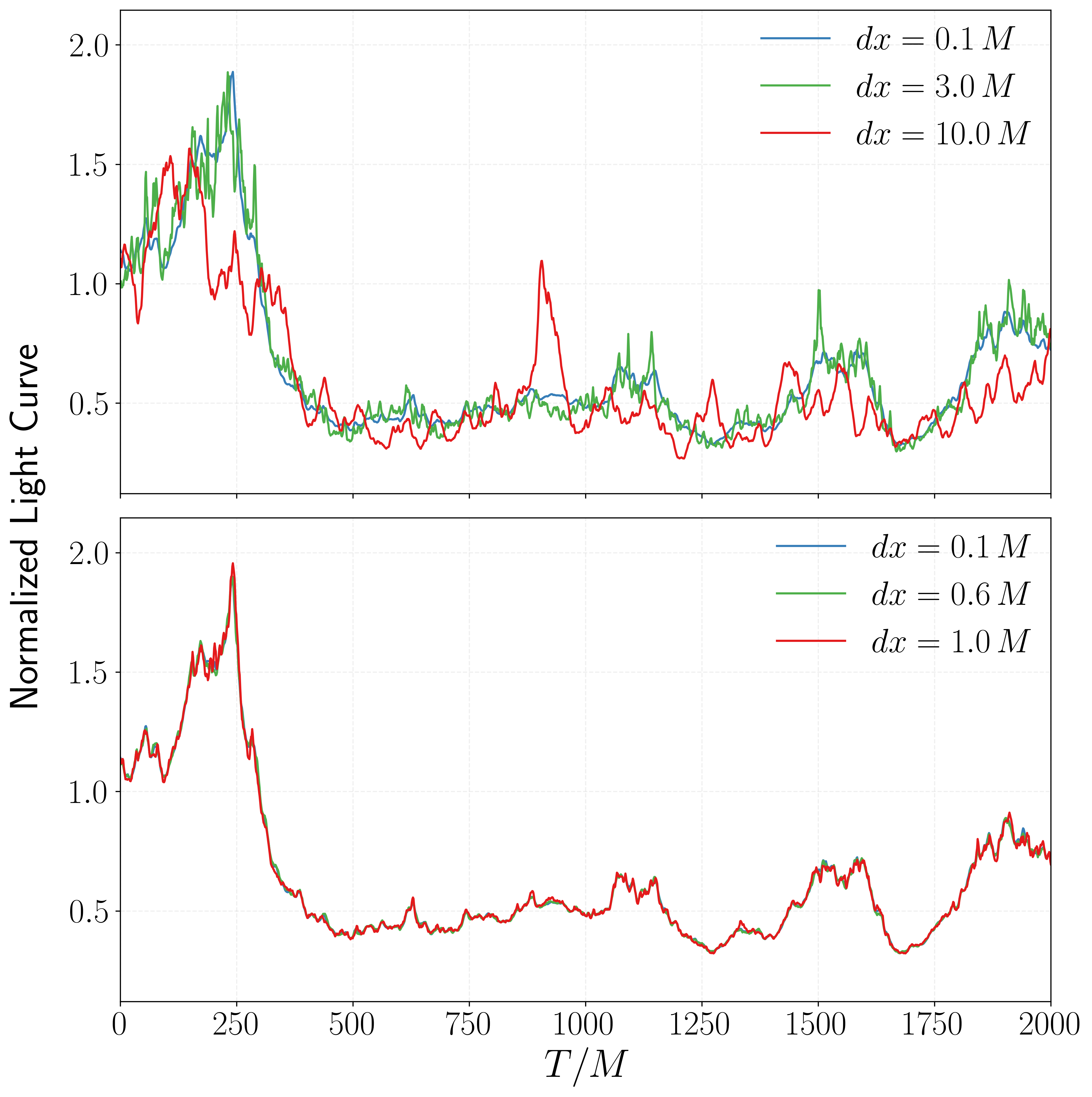}
    \caption{Same resolution test as Fig.~\ref{fig:FL_dx}, but for slow light. The same convergence threshold, $\Delta x\lesssim1M$, is recovered.}
    \label{fig:SL_dx}
\end{figure}

\begin{figure}[]
    \centering
    \includegraphics[width=\linewidth]{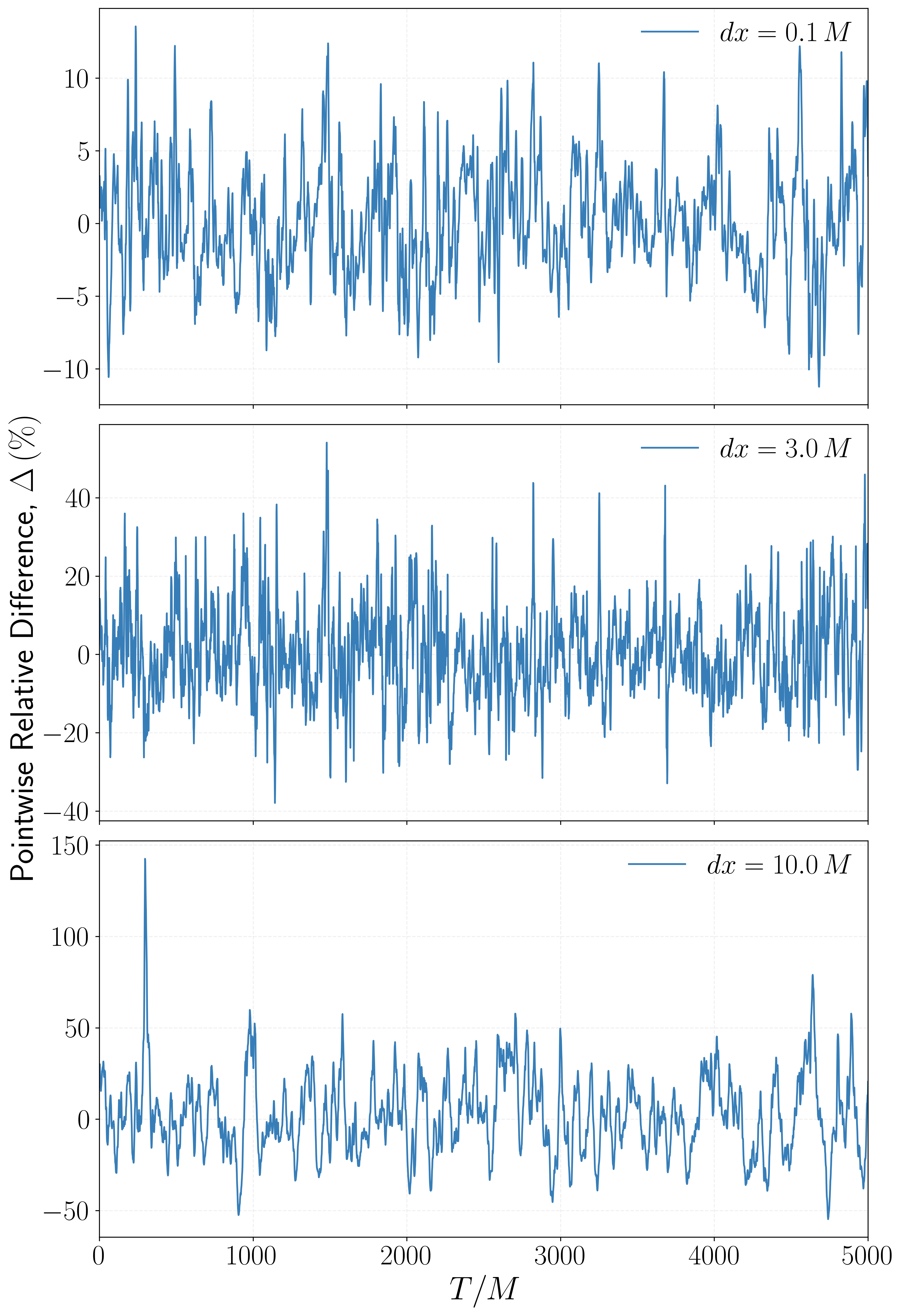}
    \caption{Pointwise relative difference between mean-normalized slow-light and fast-light light curves for several screen resolutions. Coarse grids produce large residual excursions and can mimic physical propagation effects; the fine-grid results are stable.}
    \label{fig:ER_dx}
\end{figure}

Figure~\ref{fig:ErrL1L2_dx} and Figs.~\ref{fig:FL_dx}--\ref{fig:ER_dx} summarize the spatial-resolution tests. The metrics plateau once $\Delta x\lesssim1M$, while coarser grids visibly degrade both the light curves and the fast--slow comparison. All main results use the converged regime.

\begin{figure}[]
    \centering
    \includegraphics[width=\linewidth]{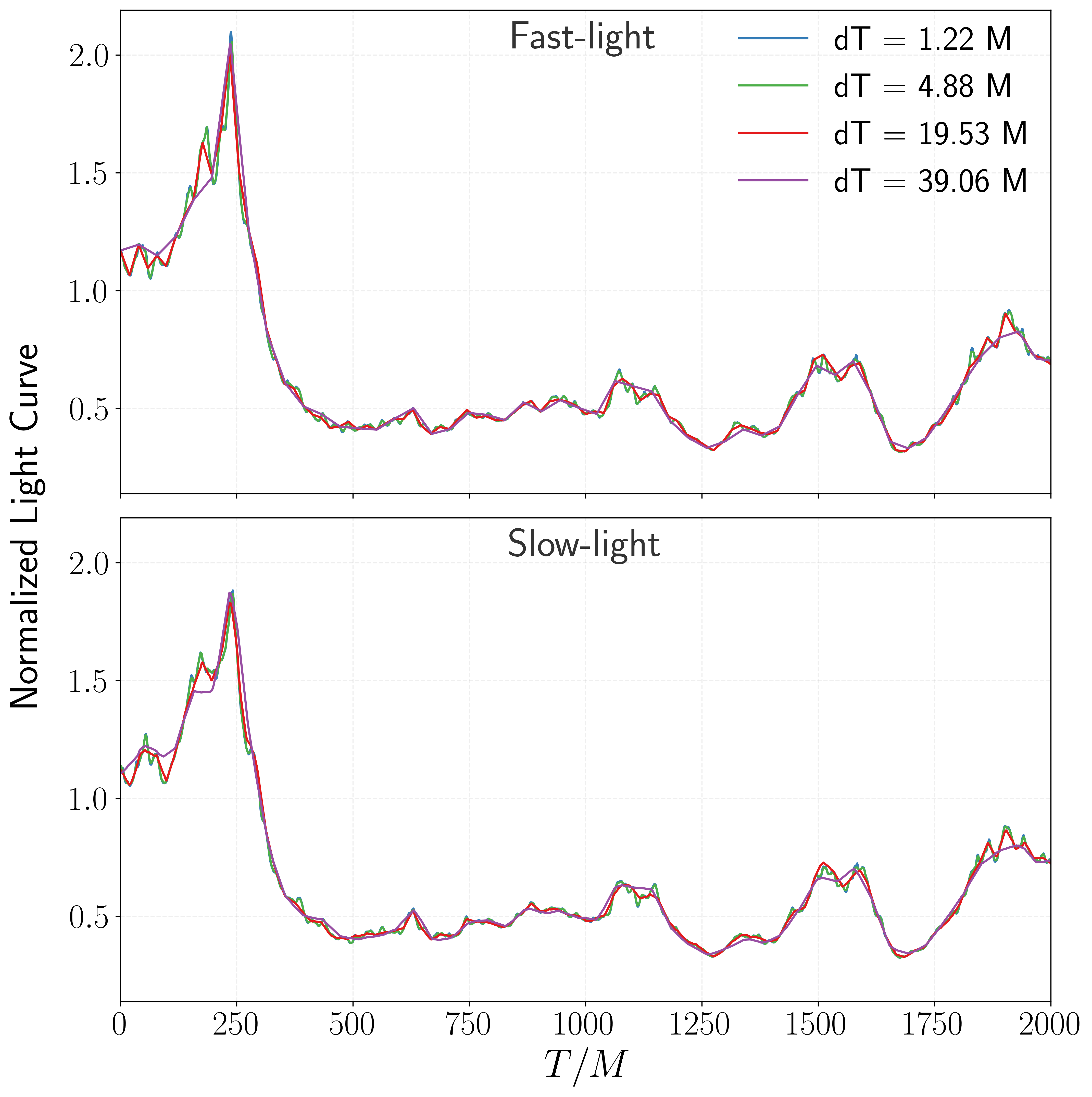}
    \caption{Mean-normalized fast-light and slow-light curves at $\theta_{\rm o}=17^\circ$ for several source cadences $\Delta T$. Coarsening the source movie suppresses high-frequency variability and makes the two prescriptions increasingly similar.}
    \label{fig:LC_dt_17}
\end{figure}

\begin{figure}[]
    \centering
    \includegraphics[width=\linewidth]{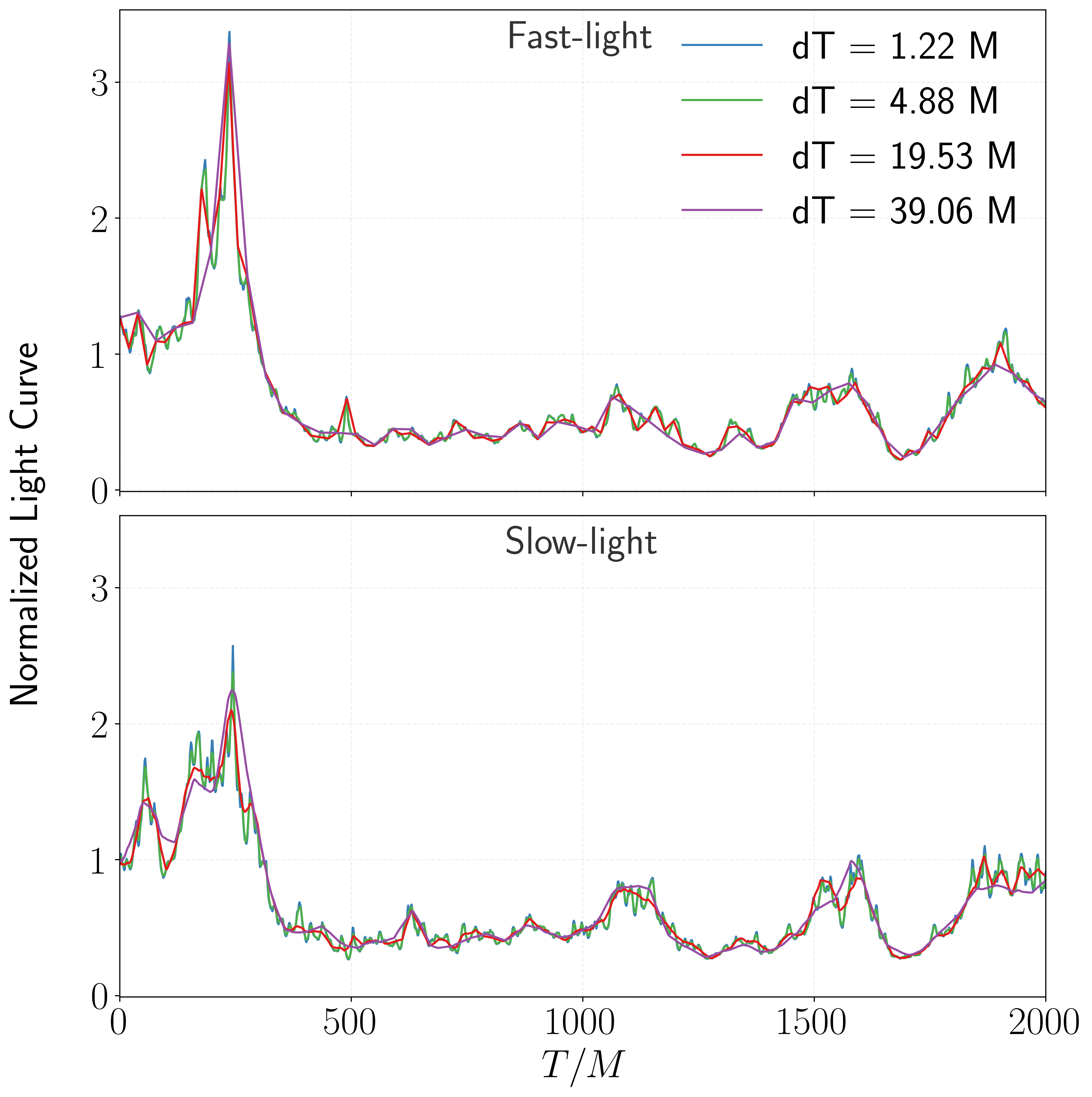}
    \caption{Same source-cadence comparison as Fig.~\ref{fig:LC_dt_17}, but for $\theta_{\rm o}=60^\circ$. At the finest cadence the separation between fast and slow light is larger than at low inclination; the difference decreases as the source becomes temporally under-resolved.}
    \label{fig:LC_dt_60}
\end{figure}

The source-cadence tests are shown in Figs.~\ref{fig:LC_dt_17} and~\ref{fig:LC_dt_60}. Both prescriptions use linear interpolation between adjacent source frames whenever the requested emission time lies between stored snapshots. Fast light queries one time per observer frame. Slow light queries the range of times selected by the screen-dependent delay map. Coarsening $\Delta T$ therefore removes short-timescale source structure and reduces the distinction between propagation prescriptions.

The temporal discretization artifacts associated with coarsening $\Delta T$ in this setup differ from the interpolation artifacts reported in GRMHD-based slow-light calculations, where nonlinear dependencies of emission and absorption coefficients on the fluid variables can generate excess power near the snapshot cadence or artificial discontinuities when the GRMHD snapshot cadence is too coarse~\cite{Wong:2024LightEchoesVLBI,Tsunetoe:2026SlowLightM87}. The \texttt{inoisy} source is a smooth Gaussian random field with controlled covariance, so the temporal discretization artifacts here have a different character. This difference reflects the source model, not the propagation prescription.

\begin{figure}[]
    \centering
    \includegraphics[width=\linewidth]{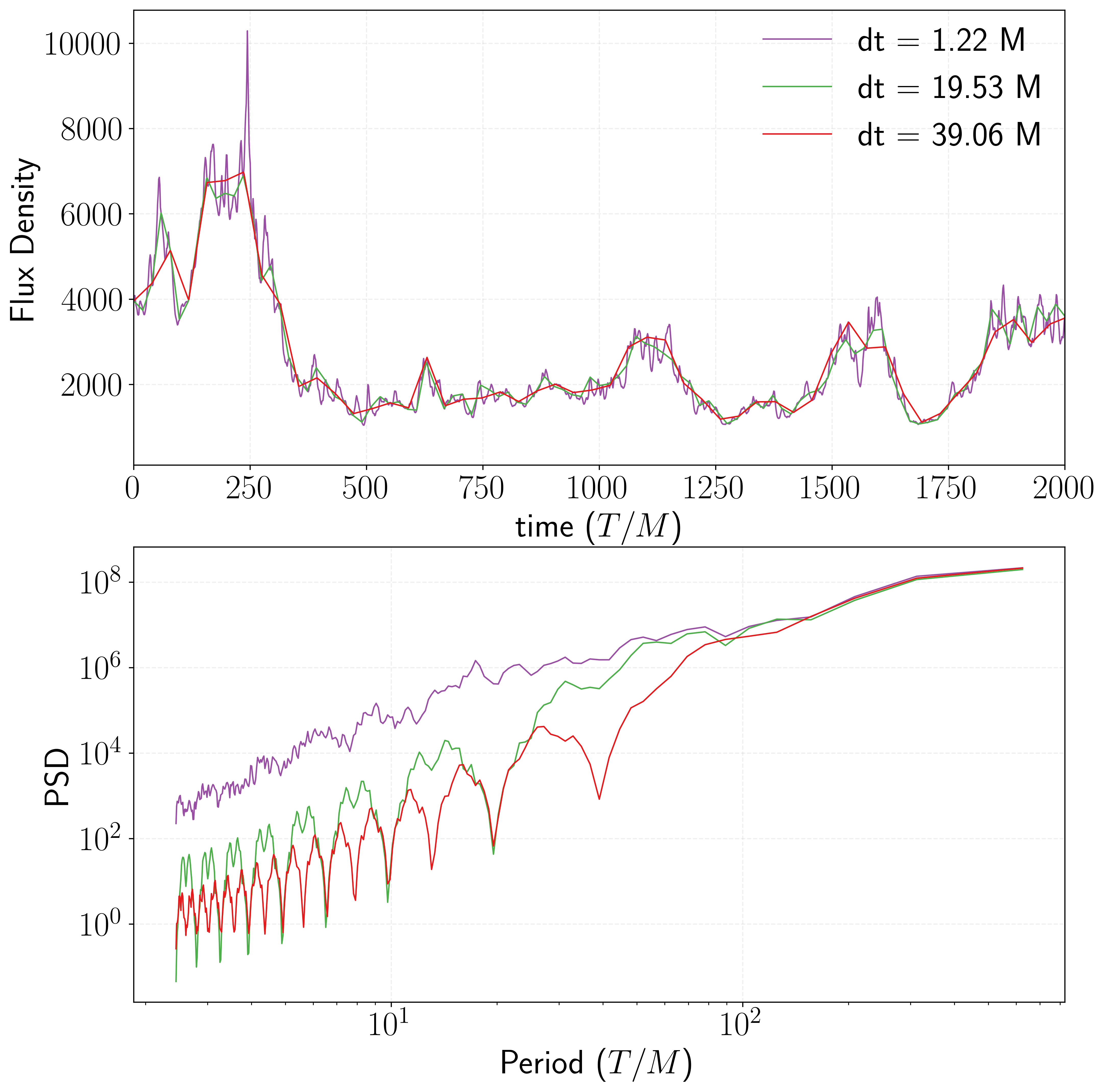}
    \caption{Effect of ray-tracing cadence on the slow-light power spectrum at $\theta_{\rm o}=60^\circ$. Top: slow-light flux-density curves for three ray-tracing cadences $\Delta t_{\rm o}$ at fixed source cadence $\Delta T$. Bottom: corresponding power spectral densities, plotted against period. The long-period spectra agree, while the coarser cadences introduce regular short-period sampling features that are absent at the finest cadence.}
    \label{fig:PSD_SlowLight}
\end{figure}
Figure~\ref{fig:PSD_SlowLight} gives a frequency-domain view of a related effect through the power spectral density (PSD), now varying the ray-tracing cadence $\Delta t_{\rm o}$ at fixed $\Delta T$. At long periods, the spectra agree across cadences. At short periods, the coarser ray-tracing cadences develop regular dips associated with the finite temporal sampling of the observed light curve. These are not physical source features. They motivate using the finest available ray-tracing cadence for the main propagation comparisons, completing the convergence checks in $\Delta x$, $\Delta T$, and $\Delta t_{\rm o}$ used throughout this work.

\clearpage

\bibliographystyle{utphys}
\bibliography{refs}

\end{document}